\documentclass[10pt,journal,twoside]{IEEEtran}
\usepackage{multirow}
\usepackage{amssymb}
\usepackage[dvips]{graphicx}
\usepackage{amsmath}
\usepackage{amsthm}
\usepackage{latexsym,bm}
\usepackage{color}
\usepackage{subfigure}
\usepackage{longtable}
\usepackage{accents}
\usepackage{cite}
\usepackage{enumerate}
\usepackage{arydshln}
\usepackage{slashbox}
\usepackage{algorithmic}
\usepackage{float}
\usepackage{booktabs}
\usepackage{subfigure}
\usepackage{stfloats}
\usepackage{graphicx}

\def\bc{{\mathbf{c}}}   
   \def\bx{{\mathbf{x}}}

\makeatletter
\def\widebar{\accentset{{\cc@style\underline{\mskip10mu}}}}
\def\Widebar{\accentset{{\cc@style\underline{\mskip8mu}}}}
\makeatother

\theoremstyle{plain}

\theoremstyle{definition}
\theoremstyle{definition}

\begin{document}

\title{{Reconfigurable Intelligent Surface-aided $M$-ary FM-DCSK System: a New Design for Noncoherent Chaos-based Communication}
\thanks{H.~Ma and Y.~Fang are with the School of Information Engineering, Guangdong University of Technology, Guangzhou 510006, China, (email: mh-zs@163.com; fangyi@gdut.edu.cn).}
\thanks{P.~Chen is with the Department of Electronic Information, Fuzhou University, Fuzhou 350116, China (e-mail: ppchen.xm@gmail.com).}
\thanks{Y.~Li is with the School of Electrical and Information Engineering, The University of Sydney, Sydney, NSW 2006, Australia (e-mail: yonghui.li@sydney.edu.au).}}
\author{\fontsize{11pt}{\baselineskip}\selectfont {Huan Ma, Yi Fang, {\em Member, IEEE}, Pingping Chen, {\em Member, IEEE}, Yonghui Li, {\em Fellow, IEEE}}\vspace{-5mm}}

\maketitle
\begin{abstract}
In this paper, we propose two reconfigurable intelligent surface-aided $M$-ary frequency-modulated differential chaos shift keying (RIS-$M$-FM-DCSK) schemes. In scheme I, the RIS is regarded as a transmitter at the source to incorporate the $M$-ary phase-shift-keying ($M$-PSK) symbols into the FM chaotic signal and to reflect the resultant $M$-ary FM chaotic signal toward the destination. The information bits of the source are carried by both the positive/negative state of the FM chaotic signal and the $M$-PSK symbols. In scheme II, the RIS is treated as a relay so that both the source and relay can simultaneously transmit their information bits to the destination. The information bits of the source and relay are carried by the positive/negative state of the FM chaotic signal and $M$-PSK symbols generated by the RIS, respectively. The proposed RIS-$M$-FM-DCSK system has an attractive advantage that it does not require channel state information for detection, thus avoiding complex channel estimation. Moreover, we derive the theoretical expressions for bit error rates (BERs) of the proposed RIS-$M$-FM-DCSK system with both scheme I and scheme II over multipath Rayleigh fading channels. Simulations results not only verify the accuracy of the theoretical derivations, but also demonstrate the superiority of the proposed system. The proposed RIS-$M$-FM-DCSK system is a promising low-cost, low-power, and high-reliability alternative for wireless communication networks.
\end{abstract}
\begin{keywords}
Reconfigurable intelligent surface (RIS), differential chaos shift keying (DCSK), channel state information, multipath Rayleigh fading channel.
\end{keywords}

\section{Introduction}\label{sect:introduction}
Chaos-based communication is considered as a potential alternative to the conventional direct sequence-spread spectrum communication\cite{Chi2003Chaos,9098915}. Thanks to natural characteristics of chaotic signal, chaos-based communication benefits from various advantages such as excellent anti-intercept, auto-correlation, non-periodic, and wideband properties\cite{9507354}. In the past three decades, differential chaos shift keying (DCSK), which is regarded as the most popular chaos-based modulation, has owned considerable development. DCSK only needs a non-coherent demodulator to avoid the chaos synchronization, which is very desired for  practical implementation in low-power and low-cost wireless-communication applications.

The theoretical bit-error-rate (BER) performance of DCSK system in different types of wireless channels has been investigated in\cite{Zhou2008Performance,1362943}. With the above theoretical advances, the DCSK system has been further designed to adapt to different scenarios, such as magnetic coupled human body communication\cite{8564572}, power line communication\cite{9204462}, underwater acoustic communication\cite{2019Multi,9761198}, ambient backscatter communication\cite{9525461} and vehicle-to-vehicle communication\cite{9732656}. To improve the transmission reliability of DCSK system, some powerful error correction codes, such as low-density parity-check codes\cite{9519519,9600574}, have been exploited to construct highly robust coded modulation systems\cite{9684235}.

In recent years, various variants of the DCSK system have been presented. To avoid the drawback of the non-constant energy in the conventional DCSK system, a frequency-modulated (FM) DCSK system has been devised in\cite{899922}. Afterwards, some ultra-wideband systems based on FM-DCSK modulation have been proposed\cite{5754623,5371909,8606201}. Although the introduction of FM can improve the performance of traditional DCSK system, it data rate and energy efficiency are low because half of the chaotic chips are used to transmit the reference signal. Therefore, designing enhanced DCSK systems with higher data rate and energy efficiency has become an urgent issue. In particular, a code-shifted DCSK system with a delay-line-free receiver has been proposed in\cite{S0218127411028829}, where the reference and information-bearing signals are sent in the same time slot. An improved code-shifted DCSK system with reference diversity has been designed in\cite{8952773}. Thanks to the reference diversity, this system has better BER performance than the conventional code-shifted DCSK system over fading channels. To make full use of frequency domain resources, some multi-carrier (MC) DCSK systems have been proposed\cite{7332774,9761226}. These MC-DCSK systems not only have the advantage of the delay-line-free structure, but also have higher data rate than the conventional DCSK system. To further improve the data rate and efficiency of the DCSK system, the joint design of the index modulation (IM) technique and the DCSK system has been proposed recently. Besides, a code-index-modulation (CIM) DCSK has been developed in\cite{7751600} and optimized in\cite{8290668}. In the CIM-DCSK system, the row index of the selected Walsh code sequence in the Walsh matrix is used to convey the information bits. In \cite{9277910,9372875}, more than one transmission entities are used for IM. A joint carrier-code IM-aided $M$-ary DCSK system was proposed in\cite{9277910}. In such a system, the carrier and the row index of the selected Walsh code sequence are used to carry extra information bits. In \cite{9372875}, a joint time-frequency-IM-aided system has been proposed, where all subcarriers and time slots are used to carry information bits.

Although various excellent DCSK variants have been proposed, how to maintain the reliability of the DCSK system in transmission scenarios with relatively long distance or some obstacles still encounters great challenges. To achieve better reliability for relatively long-distance communications, an analog space-time block coded (STBC) DCSK system was devised in\cite{6482240}. Compared with the single-antenna DCSK system, the analog STBC-DCSK system has a great improvement in terms of BER performance over multipath fading channels. An orthogonal multi-codes multiple-input-multiple-output (OMC-MIMO) MIMO-DCSK system was proposed in\cite{6661866} to further improve the BER performance. To boost the data rate of the multiple-antenna DCSK system, a multiple-input-single-output code-index-modulation-aided multiple-carrier $M$-ary DCSK (MISO-CIM-MC-$M$DCSK) system was proposed in\cite{9184069}, which inherits the merits of DCSK modulation, multiple-antenna technique, MC technique, and IM technique. To overcome the adverse effects of obstacles in transmission, some cooperative DCSK systems have been proposed, such as the amplify-and-forward relaying DCSK system\cite{iet:/content/journals/10.1049/iet-com.2014.0232} and two-way relay network-coded DCSK system\cite{8036271}. However, the relays of these systems are active, which is not conducive to the deployment in practical applications. To resolve this problem, a simultaneous wireless information and power transfer (SWIPT) cooperative DCSK system with passive relay has been proposed in\cite{9434934}. Furthermore, a buffer-aided DCSK-SWIPT system was developed in\cite{9142258}. In this system, the relay works by harvesting the energy of the received signal. Besides, two link-selection protocols have been designed for the buffer-aided DCSK-SWIPT system.

As a passive-communication-aided device, reconfigurable intelligent surface (RIS) is a promising technology\cite{9530717,9475155}. The RIS is a plane consisting of a large number of passive reflective elements, each of which can control the amplitude and/or phase of the incident signal\cite{9326394}. Based on these characteristics, the RIS has the distinctive capability to transform uncontrollable and bad channel condition into controllable and good channel condition when the channel state information (CSI) is available at the RIS\cite{9347538}. On the contrary, when the CSI is unavailable at the RIS, though the RIS cannot maximize the received SNR according to the channel phases, it can improve the received SNR using the passive reflective elements\cite{8796365,8801961}. Unlike multi-antenna and relay solutions, the RIS does not require multiple radio frequency (RF) chains, hence enabling the advantages of low power and low cost\cite{9405433}.

\begin{figure*}[!tbp]
\centering\vspace{-3mm}
\subfigure[\hspace{-0.8cm}]{ \label{fig:subfig:1a}
\includegraphics[width=7.09in,height=1.44257in]{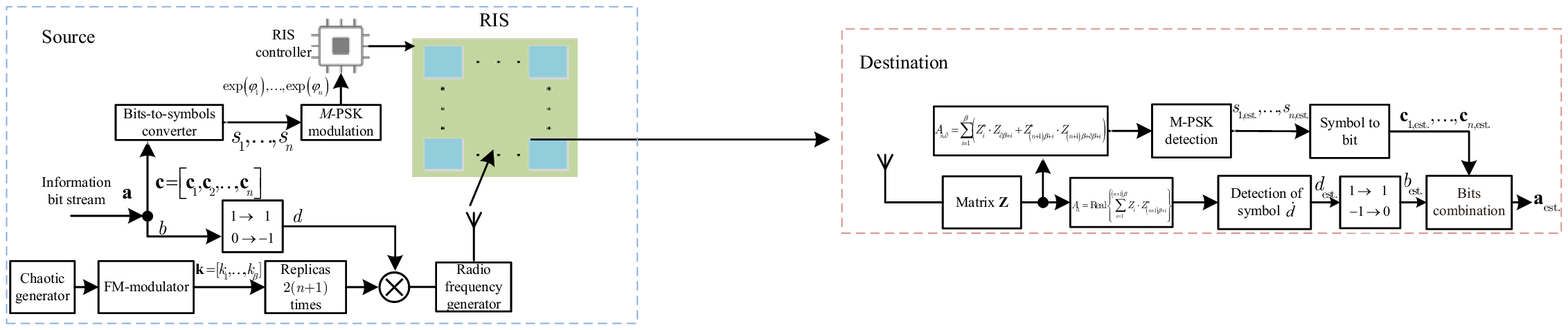}}\vspace{-1.5mm}
\subfigure[\hspace{-0.8cm}]{ \label{fig:subfig:1b}
\includegraphics[width=7.09in,height=1.7in]{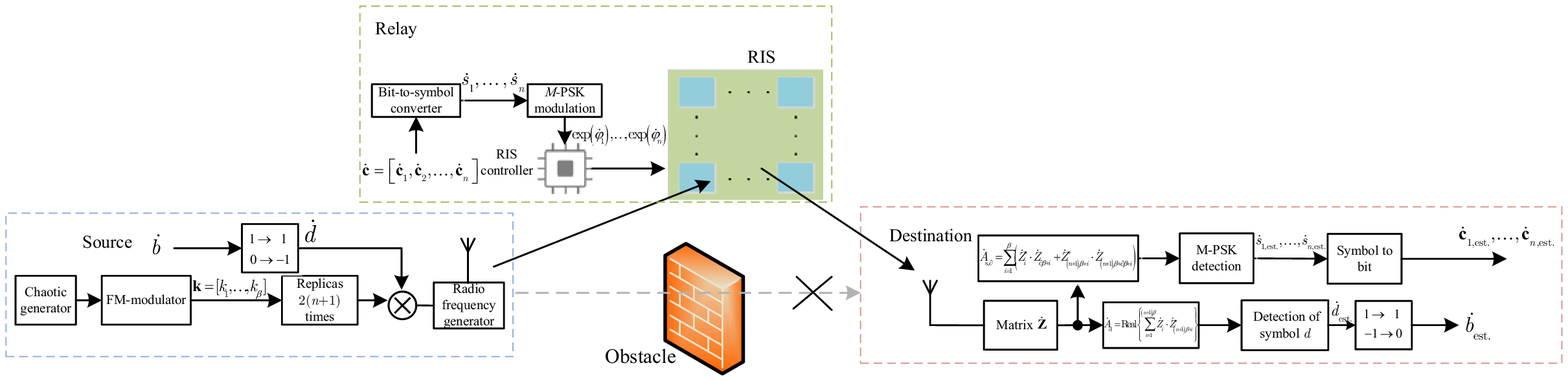}}
\vspace{-0.1cm}
\caption{Block diagrams of (a) scheme I and (b) scheme II in the proposed RIS-$M$-FM-DCSK system. }
\label{fig:fig1}  
\vspace{-2mm}
\end{figure*}
Motivated by the above advantages of RIS, we conceive an RIS-aided $M$-ary FM-DCSK system (i.e., RIS-$M$-FM-DCSK system) in this paper, which can realize high-reliability transmissions over multipath Rayleigh channels. In addition, the RIS does not require any CSI in the proposed RIS-$M$-FM-DCSK system. The contributions of this paper are summarized as follows:
\begin{enumerate}[1)]
\item \label{}
We develop two different transmission schemes, referred to as {\em scheme I and scheme II}, to improve the reliability for the proposed RIS-$M$-FM-DCSK system. Scheme I is suitable for the communication scenario (CS) that the direct link exists between the source and the destination. In this scheme, the RIS is considered as a transmitter at the source. Moreover, the information bits are not only carried by the positive/negative state of the FM chaotic signal, but also carried by the $M$-ary phase-shift-keying ($M$-PSK) symbols obtained by the phase adjustment of the RIS. Scheme II is suitable for the CS that the direct link between the source and the destination is blocked by an obstacle. In this scheme, the RIS is treated as a relay to realize the information transmissions simultaneously from the source and relay to the destination. Moreover, the information bits of the source and relay are carried by the positive/negative state of the FM chaotic signal and $M$-PSK symbols generated by the RIS, respectively.

\item \label{}
We carefully analyze the theoretical BER performance of the two proposed transmission schemes in the RIS-$M$-FM-DCSK system over multipath Rayleigh fading channels.

\item \label{}
We carry out various simulations to verify the accuracy of the proposed BER expressions. Furthermore, we illustrate that the proposed RIS-$M$-FM-DCSK system achieves significant performance gains over the existing DCSK counterparts.

\end{enumerate}

The remainder of this paper is organized as follows. The system model of the proposed RIS-$M$-FM-DCSK system is presented in section~\ref{sect:system model}. Section~\ref{sect:BER Performance Analysis} carries out the theoretical BER performance analysis of the proposed system. Section~\ref{sect:Simulation Results and Discussion} presents the simulation results and discussions. Finally, Section~\ref{sect:Conclusions} draw the conclusion.

\section{System Model}\label{sect:system model}
In this section, we describe system model of the proposed RIS-$M$-FM-DCSK modulation.
We consider two CSs: i) the direct link exists between the source and the destination; ii) the direct link between the source and the destination is blocked by an obstacle.
In the CS-i, we treat the RIS as a transmitter.
In the CS-ii, we treat the RIS as a relay.\footnote{Since RIS is treated as a relay in scheme II, ``relay'' may be used to represent ``RIS'' in scheme II from this point onwards.}
Based on the characteristics of RIS, we separately design two transmission schemes, i.e., scheme I and scheme II, for the above two CSs. The block diagrams for the two transmission schemes are illustrated in Fig.~\ref{fig:fig1}.
\subsection{RIS-M-FM-DCSK System: Scheme I} \label{sect:Scheme I in the RIS-M-FM-DCSK system}
Fig.~\ref{fig:subfig:1a} shows the system model for the proposed RIS-$M$-FM-DCSK with scheme I. In this scheme, the RIS is very close to the transmit antenna, hence we assume that the transmission between the RIS and transmit antenna is not affect by fading. As such, the RIS can be treated as a transmitter. Referring to Fig.~\ref{fig:subfig:1a}, the information bit stream ${\bf{a}} = \left[ {b,{\bf{c}}} \right]$ of the source is divided into two parts, i.e., the bit $b$ and the bit stream ${\bf{c}} = \left[ {{{\bf{c}}_1}, \ldots ,{{\bf{c}}_\partial }, \ldots ,{{\bf{c}}_n}} \right]$, where the dimension of ${{\bf{c}}_\partial }$ is $1 \times {\log _2}M$ and $1 \le \partial \le n$. Then, the bit $b$ is converted into a symbol $d = 2b - 1$, while the bit sub-streams ${\bc_1}, \ldots ,{\bc_n}$ are converted to symbols ${s_1}, \ldots ,{s_n}$, respectively, by using an $M$-ary bit-to-symbol converter. The number of the information bits carried by each transmitted symbol in scheme I equals ${\Xi _{{\rm{I}},{\rm{tot}}}} = 1 + n{\log _2}M$. To generate a signal that can carry these ${\Xi _{{\rm{I}},{\rm{tot}}}}$ information bits, the length-$\beta$ chaotic signal $\bx =[x_1, x_2,\ldots, x_\beta] $ is first modulated by using FM modulator to generate FM chaotic signal ${\bf{k}} = [{k_1}, k_2, \ldots ,{k_\beta }]$, where logistic map is employed by chaotic generator and the chip duration is defined as ${T_{\rm{c}}}$. 
Then, the chaotic signal $\bf{k}$ is repeated $2\left( {n + 1} \right)$ times to modulate the symbols $d$ and ${s_1}, \ldots ,{s_n}$. More specifically, the symbols ${s_1}, \ldots ,{s_n}$ are modulated via an $M$-PSK modulator, thus generating the phases ${\varphi _1}, \ldots ,{\varphi _n}$, where ${\varphi _\partial } \in \left( {0,2\pi } \right]$ ($1 \le \partial  \le n$).
 Meanwhile, the symbol $d$ is conveyed by the discrete baseband signal ${{\bf{T}}_{{\rm{RF}}}}$, which is sent from the transmit antenna to the RIS. In particular, ${{\bf{T}}_{{\rm{RF}}}}$ can be expressed as
\begin{align}
{{\bf{T}}_{{\rm{RF}}}} = \left[ {\underbrace {{\bf{k}},{\bf{k}}, \ldots {\bf{k}}}_{n + 1\ {\rm{ items}}},d\left( {\underbrace {{\bf{k}},{\bf{k}}, \ldots ,{\bf{k}}}_{n + 1\ {\rm{ items}}}} \right)} \right].
 \label{eq:1func}
\end{align}
The RIS reflects the incident signal with phase-related reflection coefficients $\exp \left( {j{\varphi _1}} \right), \ldots ,\exp \left( {j{\varphi _n}} \right)$, where $j = \sqrt { - 1} $. Therefore, for the $l$-th RIS reflecting element, the reflected baseband signal can be expressed as
\begin{align}
\begin{array}{l}
{{\bf{T}}_{{\rm{el}},l}} = \left[ {\underbrace {{\bf{k}},\exp \left( {j{\varphi _1}} \right){\bf{k}}, \ldots ,\exp \left( {j{\varphi _n}} \right){\bf{k}}}_{{{\bf{R}}_{\rm{d}}}},} \right.\\
\ \ \ \ \ \ \ \ \ \left. {\underbrace {d \cdot {\bf{k}},d \cdot \exp \left( {j{\varphi _1}} \right){\bf{k}}, \ldots ,d \cdot \exp \left( {j{\varphi _n}} \right){\bf{k}}}_{{{\bf{I}}_{\rm{d}}}}} \right]\\
 \ \ \ \ \ \ = \left[ {{{{T}}_{{\rm{el}},l,1}}, \ldots ,{{{T}}_{{\rm{el}},l,\Im }}, \ldots ,{{{T}}_{{\rm{el}},l,2\left( {n + 1} \right)\beta }}} \right],
\end{array}
 \label{eq:2func}
\end{align}
where $1 \le l \le N$, $N$ is the number of the RIS reflecting elements, the signals ${{\bf{R}}_{\rm{d}}}$ and ${{\bf{I}}_{\rm{d}}}$ can be regarded as the reference part and information-bearing part in the transmission of symbol $d$, respectively. The symbol $d$ is carried by the state of signal ${{\bf{I}}_{\rm{d}}}$. To be specific, the signal ${{\bf{I}}_{\rm{d}}}$ equals the positive of signal ${{\bf{R}}_{\rm{d}}}$ if $d=1$;
otherwise (i.e., $d =  - 1$), it equals the negative of signal ${{\bf{R}}_{\rm{d}}}$. Besides, the signals $\bf{k}$ (resp. $d \cdot {\bf{k}}$) and $\exp \left( {j{\varphi _\partial }} \right) {\bf{k}}$ (resp. $d \cdot \exp \left( {j{\varphi _\partial }} \right) {\bf{k}}$) in Eq.~(\ref{eq:2func}) can be regarded as reference part and information-bearing part in the transmission of symbol ${s_\partial }$ ($1 \le \partial  \le n$), respectively. In this sense, the spreading factor $SF$ of scheme I is defined as $SF = 2\left( {n + 1} \right)\beta $. Hence, the data rate of scheme I is ${R_{\rm{I}}} = \frac{{{\Xi _{{\rm{I}},{\rm{tot}}}}}}{{SF \cdot {T_{\rm{c}}}}}$.

At the destination, the received baseband signal can be written as
\begin{align}
  {r_\Im } &= \sqrt {D_{{\rm{sd}}}^{ - \varepsilon }} \sum\limits_{{\rho _{{\rm{sd}}}} = 1}^{{L_{{\rm{sd}}}}} {\sum\limits_{l = 1}^N {{\alpha _{{\rm{sd,}}l{{,}}{\rho _{{\rm{sd}}}}}}{T_{{\rm{el}},l,\Im  - {\tau _{l,{\rho _{{\rm{sd}}}}}}}} + {w_\Im }} }  \hfill \nonumber\\
   &= \sqrt {D_{{\rm{sd}}}^{ - \varepsilon }} \sum\limits_{{\rho _{{\text{sd}}}} = 1}^{{L_{{\rm{sd}}}}} {{\Lambda _{{\rm{sd}},{\rho _{{\rm{sd}}}}}}{T_{{\rm{el}},1,\Im  - {\tau _{{\rho _{{\rm{sd}}}}}}}} + {w_\Im }} , \hfill
   \label{eq:3func}
\end{align}
where ${D_{{\rm{sd}}}}$ is the distance between the RIS and destination, $\varepsilon $ is the path loss coefficient\cite{John1983Digital}, ${L_{{\rm{sd}}}}$ is the number of the paths between the RIS and destination, ${\alpha _{{\rm{sd}},l,{\rho _{{\rm{sd}}}}}}$ and ${\tau _{l,{\rho _{{\rm{sd}}}}}}$ denote channel coefficient and delay of the ${\rho _{{\rm{sd}}}}$-th ($1 \le {\rho _{{\rm{sd}}}} \le {L_{{\rm{sd}}}}$) path between the $l$-th RIS reflecting element and destination, respectively, $1 \le \Im  \le 2\left( {n + 1} \right)\beta $, and ${\Lambda _{{\rm{sd}},{\rho _{{\rm{sd}}}}}} = \sum\nolimits_{l = 1}^N {{\alpha _{{\rm{sd,}}l{{,}}{\rho _{{\rm{sd}}}}}}} $. In Eq.~(\ref{eq:3func}), ${w_\Im } \sim {\cal{CN}}\left( {0,{N_0}} \right)$ is complex additive white Gaussian noise (AWGN), and ${\cal{CN}}\left( {{\phi  _1},{\phi _2}} \right)$ denotes the circularly symmetric complex Gaussian distribution with mean of ${\phi _1}$ and variance of ${\phi _2}$. In this paper, we assume that the path delays between each RIS reflecting element and destination are the same \cite{6661866,1123456,7908989,4469995}, i.e., ${\tau _{{l_1},{\rho _{{\rm{sd}}}}}} = {\tau _{{l_2},{\rho _{{\rm{sd}}}}}}$ (${l_1} = 1, \ldots ,N$, ${l_2} = 1, \ldots ,N$, and ${l_1} \ne {l_2}$). Hence, ${\tau _{l,{\rho _{{\rm{sd}}}}}}$ can be simplified to ${\tau _{{\rho _{{\rm{sd}}}}}}$. Subsequently, the received baseband signals are stored in a matrix ${\bf{Z}} = \left[ {{Z_1},{Z_2}, \ldots ,{Z_{2\left( {n + 1} \right)\beta }}} \right]$. To perform the demodulation for the symbol $d$ of the source, the decision metric is first formulated as
\begin{align}
  {A_{\rm{d}}} &= {\rm{Real}}\!\left\{ {\sum\limits_{i = 1}^{\left( {n + 1} \right)\beta } {{Z_i} \cdot Z_{\left( {n + 1} \right)\beta  + i}^ * } } \right\} \hfill \nonumber\\
   &\approx {\rm{Real}}\left\{ {\sum\limits_{{i_1} = 1}^{n + 1} {\sum\limits_{{i_2} = 1}^\beta  {\left( {\sqrt {D_{{\text{sd}}}^{ - \varepsilon }} \sum\limits_{{\rho _{{\rm{sd}}}} = 1}^{{L_{{\rm{sd}}}}} {{\Lambda _{{\rm{sd}},{\rho _{{\text{sd}}}}}}\exp \left( {j{\varphi _{{i_1} - 1}}} \right){k_{{i_2}}}} } \right.} } } \right. \hfill \nonumber\\
  &\ \ \left. { \!+ {w_{\left( {{i_1} - 1} \right)\beta \! +\! {i_2}}}} \right)\!\! \times\!\! \left( {d \!\cdot\! \sqrt {D_{{\rm{sd}}}^{ - \varepsilon }}\! \sum\limits_{{\rho _{{\rm{sd}}}} = 1}^{{L_{{\rm{sd}}}}} {{\Lambda _{{\rm{sd}},{\rho _{{\rm{sd}}}}}}\exp \left( {j{\varphi _{{i_1} - 1}}} \right){k_{{i_2}}}} } \right. \hfill \nonumber\\
  &\ \left. {{{\left. { + {w_{\left( {n + 1} \right)\beta  + \left( {{i_1} - 1} \right)\beta  + {i_2}}}} \right)}^*}} \right\}, \hfill
  \label{eq:4func}
\end{align}
where ${\rm{Real}}\left\{  \cdot  \right\}$ is the real part of the complex variable, $ * $ is conjugation operator and ${\varphi _0} = 0$. Based on ${A_{\rm{d}}}$, the symbol $d$ can be estimated as
\begin{align}
{d_{{\rm{est}}}} = \left\{ \begin{gathered}
  1,\ \ \ \ {A_{\rm{d}}} \geqslant 0 \hfill \\
   -1,\ \ {\text{Otherwise}} \hfill \\
\end{gathered}  \right..
  \label{eq:5func}
\end{align}
Then, the estimated symbol ${d_{{\rm{est}}}}$ is converted into the estimated bit ${b_{{\rm{est}}}}$. Specifically, ${b_{{\rm{est}}}} = 1$ when ${d_{{\rm{est}}}} = 1$, otherwise, ${b_{{\rm{est}}}} = 0$.

On the other hand, to perform the demodulation for the symbol ${s_\partial }$ ($1 \leqslant \partial  \leqslant n$) of the source, the decision metric is first computed as
\begin{align}
  {A_{{\rm{s}},\partial }} &= \sum\limits_{i = 1}^\beta  {\left( {Z_i^ *  \cdot {Z_{\partial \beta  + i}} + Z_{\left( {n + 1} \right)\beta  + i}^ *  \cdot {Z_{\left( {n + 1} \right)\beta  + \partial \beta  + i}}} \right)}  \hfill \nonumber\\
   &\approx \sum\limits_{i = 1}^\beta  {\left[ {{{\left( {\sqrt {D_{{\rm{sd}}}^{ - \varepsilon }} \sum\limits_{{\rho _{{\rm{sd}}}} = 1}^{{L_{{\rm{sd}}}}} {{\Lambda _{{\rm{sd}},{\rho _{{\rm{sd}}}}}}{k_i} + {w_i}} } \right)}^ * }} \right.}  \hfill \nonumber\\
   &\times \left( {\sqrt {D_{{\rm{sd}}}^{ - \varepsilon }} \sum\limits_{{\rho _{{\rm{sd}}}} = 1}^{{L_{{\rm{sd}}}}} {{\Lambda _{{\rm{sd}},{\rho _{{\rm{sd}}}}}}\exp \left( {j{\varphi _\partial }} \right){k_i} + {w_{\partial \beta  + i}}} } \right) \hfill \nonumber\\
   &+ {d^2}{\left( {\sqrt {D_{{\rm{sd}}}^{ - \varepsilon }} \sum\limits_{{\rho _{{\rm{sd}}}} = 1}^{{L_{{\rm{sd}}}}} {{\Lambda _{{\rm{sd}},{\rho _{{\rm{sd}}}}}}{k_i} + {w_{\left( {n + 1} \right)\beta  + i}}} } \right)^ * } \hfill \nonumber\\
  &\left. { \times\!\! \left( \!{\sqrt {D_{{\rm{sd}}}^{ - \varepsilon }} \sum\limits_{{\rho _{{\rm{sd}}}} = 1}^{{L_{{\rm{sd}}}}} {{\Lambda _{{\rm{sd}},{\rho _{{\rm{sd}}}}}}\exp \left( {j{\varphi _\partial }} \right){k_i} \!+\! {w_{\left( {n + 1} \right)\beta  + \partial \beta  + i}}} } \!\right)} \right]\!. \hfill
  \label{eq:6func}
\end{align}
Hence, based on ${A_{{\rm{s}},\partial }}$, the estimated symbol ${s_{\partial ,{\rm{est}}}}$ can be obtained by using $M$-PSK demodulation. Then the estimated bit sub-stream ${{\bf{c}}_{\partial ,{\rm{est}}}}$ can be obtained from an $M$-ary symbol-to-bit converter.
\subsection{RIS-M-FM-DCSK System: Scheme II} \label{sect:Scheme II in the RIS-M-FM-DCSK system}
Fig.~\ref{fig:subfig:1b} shows the system model for the RIS-$M$-FM-DCSK with scheme II. In the CS-ii, the source communicates with the destination through the RIS, which is treated as a relay.
As shown in Fig.~\ref{fig:subfig:1b}, the information bit of the source is $\dot b$ and information bit stream of the relay is $\dot {\bf{c}} = \left[ {{{\dot {\bf{c}}}_1}, \ldots ,{{\dot {\bf{c}}}_\partial }, \ldots ,{{\dot {\bf{c}}}_n}} \right]$, where the dimension of ${\dot {\bf{c}}_\partial }$ is $1 \times {\log _2}M$. Then, the bit $b$ is converted into a symbol $\dot d = 2\dot b - 1$, while the bit sub-streams ${\dot {\bf{c}}_1}, \ldots ,{\dot {\bf{c}}_n}$ are converted to symbols ${\dot s_1}, \ldots {\dot s_n}$ by using an $M$-ary bit-to-symbol converter. In each transmission period of scheme II, the number of the information bits sent from the source to the destination is ${\Xi _{{\rm{II}},{\rm{s}}}} = 1$, while the number of the information bits sent from the relay to the destination is ${\Xi _{{\rm{II}},{\text{r}}}} = n{\log _2}M$.

Next, at the source, the symbol $\dot d$ is conveyed by the discrete baseband signal ${\dot {\bf{T}}_{{\rm{RF}}}}$, which is emitted from the antenna towards the RIS. Specifically, ${\dot {\bf{T}}_{{\rm{RF}}}}$ is expressed as
\begin{align}
{\dot {\bf{T}}_{{\rm{RF}}}} = \left[ {\underbrace {{\bf{k}},{\bf{k}}, \ldots {\bf{k}}}_{n + 1\ {\text{ items}}},\dot d\left( {\underbrace {{\bf{k}},{\bf{k}}, \ldots ,{\bf{k}}}_{n + 1\ {\text{ items}}}} \right)} \right].
\label{eq:7func}
\end{align}
At the relay, the symbols ${\dot s_1}, \ldots {\dot s_n}$ are modulated via an $M$-PSK modulator, thus generating the phases ${\dot \varphi _1}, \ldots ,{\dot \varphi _n}$, where ${\dot \varphi _\partial } \in \left( {0,2\pi } \right]$ ($1 \leqslant \partial  \leqslant n$).
The transmitted symbols of the relay are conveyed by the reflected signals of the RIS. In particular, the reflected signal of the $l$-th RIS reflecting element is expressed as
\begin{align}
  {{\dot {\bf{T}}}_{{\rm{el}},l}} &= \left[ {\underbrace {{\bf{k}},\exp \left( {j{{\dot \varphi }_1}} \right){\bf{k}}, \ldots ,\exp \left( {j{{\dot \varphi }_n}} \right){\bf{k}}}_{{{\dot {\bf{R}}}_{\rm{d}}}},} \right. \hfill \nonumber\\
  &\ \ \ \left. {\underbrace {\dot d \cdot {\bf{k}},\dot d \cdot \exp \left( {j{{\dot \varphi }_1}} \right){\bf{k}}, \ldots ,\dot d \cdot \exp \left( {j{{\dot \varphi }_n}} \right){\bf{k}}}_{{{\dot {\bf{I}}}_{\rm{d}}}}} \right] \hfill \nonumber\\
   &= \left[ {{{\dot {{T}}}_{{\rm{el}},l,1}}, \ldots ,{{\dot {{T}}}_{{\rm{el}},l,\Im }}, \ldots ,{{\dot {{T}}}_{{\rm{el}},l,2\left( {n + 1} \right)\beta }}} \right], \hfill
   \label{eq:8func}
\end{align}
where the signals ${\dot {\bf{R}}_{\rm{d}}}$ and ${\dot {\bf{I}}_{\rm{d}}}$ can be treated as the reference part and information-bearing part in the transmission of symbol $\dot d$, respectively. The symbol $\dot d$ is conveyed by the state of signal ${\dot {\bf{I}}_{\rm{d}}}$.
To be specific, the signal ${\dot {\bf{I}}_{\rm{d}}}$ equals the positive of signal ${\dot {\bf{R}}_{\rm{d}}}$ if $\dot d=1$;
otherwise (i.e., $\dot d =  - 1$), it equals the negative of signal ${\dot {\bf{R}}_{\rm{d}}}$.
In addition, the signals ${\bf{k}}$ (resp. $\dot d \cdot {\bf{k}}$) and $\exp \left( {j{{\dot \varphi }_\partial }} \right){\bf{k}}$ (resp. $\dot d \cdot \exp \left( {j{{\dot \varphi }_\partial }} \right) {\bf{k}}$) in Eq.~(\ref{eq:8func}) can be treated as reference part and information-bearing part in the transmission of symbol ${\dot s_\partial }$ ($1 \leqslant \partial  \leqslant n$), respectively. The spreading factor $SF$ of scheme II is defined as $2\left( {n + 1} \right)\beta $. Accordingly, the data rate of the source and relay are ${R_{{\rm{II}},{\rm{s}}}} = \frac{{{\Xi _{{\rm{II}},{\rm{s}}}}}}{{SF \cdot {T_{\rm{c}}}}}$ and ${R_{{\rm{II}},{\rm{r}}}} = \frac{{{\Xi _{{\rm{II}},{\rm{r}}}}}}{{SF \cdot {T_{\rm{c}}}}}$, respectively.

At the destination, the received baseband signal can be expressed as
\begin{align}
{\dot r_\Im } &\!= \!\sqrt {D_{{\rm{sr}}}^{ - \varepsilon }\!\!D_{{\rm{rd}}}^{ - \varepsilon }}\!\! \sum\limits_{{\rho _{{\rm{sr}}}} = 1}^{{L_{{\rm{sr}}}}} \!{\sum\limits_{{\rho _{{\rm{rd}}}} = 1}^{{L_{{\rm{rd}}}}}\! {\sum\limits_{l = 1}^N {{\alpha _{{\rm{sr,}}l{\!{,}}{\rho _{{\rm{sr}}}}}}\!{\alpha _{{\rm{rd,}}l{{,}}{\rho _{{\rm{rd}}}}}}{{\dot T}_{{\rm{el}},l,\Im \! - \!{\tau _{l,{\rho _{{\rm{sr}}}}}} \!-\! {\tau _{l,{\rho _{{\rm{rd}}}}}}}} } } }\nonumber\\
 &\ \ \ + {w_\Im } ,
   \label{eq:9func}
\end{align}
where ${D_{{\rm{sr}}}}$ and ${D_{{\rm{rd}}}}$ are the distances between the source and relay and relay and destination, respectively; ${L_{{\rm{sr}}}}$ and ${L_{{\rm{rd}}}}$ are the numbers of the paths between the source and $l$-th RIS reflecting element and the $l$-th RIS reflecting element and destination, respectively; ${\alpha _{{\rm{sr}},l,{\rho _{{\rm{sr}}}}}}$ and ${\tau _{l,{\rho _{{\rm{sr}}}}}}$ denote channel coefficient and delay of the ${\rho _{{\rm{sr}}}}$-th ($1 \le {\rho _{{\rm{sr}}}} \le {L_{{\rm{sr}}}}$) path between the source and the $l$-th RIS reflecting element, respectively; ${\alpha _{{\rm{rd}},l,{\rho _{{\rm{rd}}}}}}$ and ${\tau _{l,{\rho _{{\rm{rd}}}}}}$ denote channel coefficient and delay of the ${\rho _{{\rm{rd}}}}$-th ($1 \le {\rho _{{\rm{rd}}}} \le {L_{{\rm{rd}}}}$) path between the $l$-th RIS reflecting element and the destination, respectively. For convenience, a cascaded source-relay (RIS)-destination channel via the $l$-th RIS reflecting element is modeled here.
We define the number of the paths for the source-$l$-th RIS reflecting element channel and the $l$-th RIS reflecting element-destination channel as ${L_{{\rm{sr}}}}$ and ${L_{{\rm{rd}}}}$, respectively. Thereby, the cascaded source-RIS (relay)-destination channel via the $l$-th RIS reflecting element can be modeled as a ${L_{{\rm{sr}}}}{L_{{\rm{rd}}}}$-path fading channel. The ${\rho _{{\rm{ca}}}}$-th ($1 \leqslant {\rho _{{\rm{ca}}}} \leqslant {L_{{\rm{sr}}}}{L_{{\rm{rd}}}}$) path with the channel coefficient ${\Gamma _{{\rm{sd,}}l{{,}}{\rho _{{\rm{ca}}}}}}$ is constructed by concatenating the ${\rho _{{\rm{sd}}}}$-th path between the source and the $l$-th RIS reflecting element and the ${\rho _{{\rm{sr}}}}$-th path between the $l$-th RIS reflecting element and destination. The time delay of ${\rho _{{\rm{ca}}}}$-th path at the destination is ${\tau _{l,{\rho _{{\rm{ca}}}}}} = {\tau _{l,{\rho _{{\rm{sr}}}}}} + {\tau _{l,{\rho _{{\rm{rd}}}}}}$. In this paper, we consider that the path delays of each RIS reflecting element are the same, i.e., ${\tau _{{l_1},{\rho _{{\rm{ca}}}}}} = {\tau _{{l_2},{\rho _{{\rm{ca}}}}}}$ (${l_1} = 1, \ldots ,N$, ${l_2} = 1, \ldots ,N$, and ${l_1} \ne {l_2}$). Hence, ${\tau _{l,{\rho _{{\rm{ca}}}}}}$ can be simplified to ${\tau _{{\rho _{{\rm{ca}}}}}}$.

Based on the above foundations, the received baseband signal can be rewritten as
\begin{align}
{\dot r_\Im } = \sqrt {\frac{1}{{D_{{\rm{sr}}}^\varepsilon D_{{\rm{rd}}}^\varepsilon }}} \sum\limits_{{\rho _{{\rm{ca}}}} = 1}^{{L_{{\rm{tot}}}}} {{\Gamma _{{\rm{sd},}{\rho _{{\rm{ca}}}}}}{{\dot T}_{{\rm{el}},1,\Im  - {\tau _{l,{\rho _{{\rm{ca}}}}}} }} + {w_\Im }} ,
\label{eq:10func}
\end{align}
where $\sum\limits_{{\rho _{{\rm{ca}}}} = 1}^{{L_{{\rm{tot}}}}} {{\Gamma _{{\rm{sd,}}{\rho _{{\rm{ca}}}}}}} $ is calculated as
\begin{align}
  \sum\limits_{{\rho _{{\rm{ca}}}} = 1}^{{L_{{\rm{tot}}}}} {{\Gamma _{{\rm{sd},}{\rho _{{\rm{ca}}}}}}}  &= \sum\limits_{{\rho _{{\rm{sr}}}} = 1}^{{L_{{\rm{sr}}}}} {\sum\limits_{{\rho _{{\rm{rd}}}} = 1}^{{L_{{\rm{rd}}}}} {\sum\limits_{l = 1}^N {{\alpha _{{\rm{sr}},l{{,}}{\rho _{{\rm{sr}}}}}}{\alpha _{{\rm{rd}},l{{,}}{\rho _{{\rm{rd}}}}}}} } }  \hfill \nonumber\\
   &= \sum\limits_{{\rho _{{\rm{ca}}}} = 1}^{{L_{{\rm{tot}}}}} {\sum\limits_{l = 1}^N {{\Gamma _{{\rm{sd}},l{{,}}{\rho _{{\rm{ca}}}}}}} } . \hfill
   \label{eq:11func}
\end{align}
Then, the received baseband signal ${\dot r_\Im }$ is stored in a matrix $\dot {\bf{Z}} = \left[ {{{\dot Z}_1},{{\dot Z}_2}, \ldots ,{{\dot Z}_{2\left( {n + 1} \right)\beta }}} \right]$.
To detect the symbol $\dot d$ of the source, the decision metric is first computed as
\begin{align}
  {{\dot A}_{\rm{d}}} &\!\!= {\rm{Real}}\left\{ {\sum\limits_{i = 1}^{\left( {n + 1} \right)\beta } {{{\dot Z}_i} \cdot \dot Z_{\left( {n + 1} \right)\beta  + i}^ * } } \right\} \hfill \nonumber\\
   &\approx\! {\rm{Real}}\!\left\{\! {\sum\limits_{{i_1} = 1}^{n + 1} {\sum\limits_{{i_2} = 1}^\beta \!\! {\left(\!\! {\sqrt {D_{{\rm{sr}}}^{ - \varepsilon }D_{{\rm{rd}}}^{ - \varepsilon }} \sum\limits_{{\rho _{{\rm{ca}}}} = 1}^{{L_{{\rm{tot}}}}}\! \!{{\Gamma _{{\rm{sd}},{\rho _{{\rm{ca}}}}}}\!\exp \left( {j{{\dot \varphi }_{{i_1} - 1}}} \right)\!\!{k_{{i_2}}}} } \right.} } } \right. \hfill \nonumber\\
  &\left. { + {w_{\left( {{i_1} \!-\! 1} \right)\beta  + {i_2}}}} \right) \!\!\times \!\!\left( \!\!{\dot d\sqrt {D_{{\rm{sr}}}^{ - \varepsilon }\!D_{{\rm{rd}}}^{ - \varepsilon }} \sum\limits_{{\rho _{{\rm{ca}}}} = 1}^{{L_{{\rm{tot}}}}} \!\!{{\Gamma _{{\rm{sd}},{\rho _{{\rm{ca}}}}}}\exp \left( {j{{\dot \varphi }_{{i_1} - 1}}} \right)\!\!{k_{{i_2}}}} } \right. \hfill \nonumber\\
  &\left. {{{\left. { + {w_{\left( {n + 1} \right)\beta  + \left( {{i_1} - 1} \right)\beta  + {i_2}}}} \right)}^*}} \right\}, \hfill
  \label{eq:12func}
\end{align}
where ${\dot \varphi _0} = 0$. Based on $ {{\dot A}_{\rm{d}}}$, the symbol $\dot d$ can be estimated as
\begin{align}
{\dot d_{{\rm{est}}}} = \left\{ \begin{gathered}
  \ \ 1,\ \ \ {{\dot A}_{\rm{d}}} \geqslant 0 \hfill \\
   - 1,\ \ {\text{Otherwise}} \hfill
\end{gathered}  \right.,
\label{eq:13func}
\end{align}
Then, the estimated symbol ${\dot d_{{\rm{est}}}}$ is converted into the estimated bit ${\dot b_{{\rm{est}}}}$. To be specific, ${\dot b_{{\rm{est}}}} = 1$ when ${\dot d_{{\rm{est}}}} = 1$, otherwise, ${\dot b_{{\rm{est}}}} = 0$.

To detect the symbol ${s_\partial }$ ($1 \leqslant \partial  \leqslant n$) of the source, the decision metric is first calculated as
\begin{align}
  {{\dot A}_{{\text{s}},\partial }} &= \sum\limits_{i = 1}^\beta  {\left( {\dot Z_i^ *  \cdot {{\dot Z}_{\partial \beta  + i}} + \dot Z_{\left( {n + 1} \right)\beta  + i}^ *  \cdot {{\dot Z}_{\left( {n + 1} \right)\beta  + \partial \beta  + i}}} \right)}  \hfill \nonumber\\
   &\approx \sum\limits_{i = 1}^\beta  {\left[ {{{\left( {\sqrt {D_{{\rm{sr}}}^{ - \varepsilon }D_{{\rm{rd}}}^{ - \varepsilon }} \sum\limits_{{\rho _{{\rm{ca}}}} = 1}^{{L_{{\rm{tot}}}}} {{\Gamma _{{\rm{sd}},{\rho _{{\rm{ca}}}}}}{k_i} + {w_i}} } \right)}^ * }} \right.}  \hfill \nonumber\\
   &\times \left( {\sqrt {D_{{\rm{sr}}}^{ - \varepsilon }D_{{\rm{rd}}}^{ - \varepsilon }} \sum\limits_{{\rho _{{\rm{ca}}}} = 1}^{{L_{{\rm{tot}}}}} {{\Gamma _{{\rm{sd,}}{\rho _{{\rm{ca}}}}}}\exp \left( {j{{\dot \varphi }_\partial }} \right){k_i} + {w_{\partial \beta  + i}}} } \right) \hfill \nonumber\\
   &+ {d^2}{\left( {\sqrt {D_{{\rm{sr}}}^{ - \varepsilon }D_{{\rm{rd}}}^{ - \varepsilon }} \sum\limits_{{\rho _{{\rm{ca}}}} = 1}^{{L_{{\rm{tot}}}}} {{\Gamma _{{\rm{sd,}}{\rho _{{\rm{ca}}}}}}{k_i} + {w_{\left( {n + 1} \right)\beta  + i}}} } \right)^ * } \hfill \nonumber\\
  &\left. { \times\!\! \left(\!\!\! {\sqrt {D_{{\rm{sr}}}^{ - \varepsilon }\!\!D_{{\rm{rd}}}^{ - \varepsilon }}\! \sum\limits_{{\rho _{{\rm{ca}}}} = 1}^{{L_{{\rm{tot}}}}}\!\! {{\Gamma _{{\rm{sd,}}{\rho _{{\rm{ca}}}}}}\!\exp \left( {j{{\dot \varphi }_\partial }} \right){k_i} \!\!+\! \!{w_{\left(\! {n \!+ \!1} \!\right)\beta  + \partial \beta  + i}}} }\! \right)} \!\right]\!. \hfill
  \label{eq:14func}
\end{align}
Hence, based on ${\dot A_{{\rm{s}},\partial }}$, the estimated symbol ${\dot s_{\partial ,{\rm{est}}}}$ can be obtained by using $M$-PSK demodulation. Then the estimated bit sub-stream ${\dot {\bf{c}}_{\partial ,{\rm{est}}}}$ can be obtained from an $M$-ary symbol-to-bit converter.

\subsection{Channel Model} \label{sect:Channel Model}
In this paper, we consider a multipath Rayleigh fading channel, which is commonly utilized in spread-spectrum chaotic communication systems\cite{7332774,9761226,9277910,9372875}. In addition, we assume that the channel coefficients ${\alpha _{{\rm{sd}},{l_1},{\rho _{{\rm{sd}}}}}}$, ${\alpha _{{\rm{sr}},{l_1}{{,}}{\rho _{{\rm{sr}}}}}}$, and ${\alpha _{{\rm{rd}},{l_1},{\rho _{{\rm{rd}}}}}}$ are independent of ${\alpha _{{\rm{sd}},{l_2},{\rho _{{\rm{sd}}}}}}$, ${\alpha _{{\rm{sr}},{l_2}{{,}}{\rho _{{\rm{sr}}}}}}$, and ${\alpha _{{\rm{rd}},{l_2},{\rho _{{\rm{rd}}}}}}$, respectively\cite{9326394,8796365}.
According to the central limit theorem\cite{8796365,8801961}, the random variables ${\Lambda _{{\rm{sd}},{\rho _{{\rm{sd}}}}}}$ and ${\Gamma _{{\rm{sd}},{\rho _{{\rm{ca}}}}}}$ follow the complex Gaussian distributions ${\cal{CN}}\left( {0,N{\rm{E}}\left\{ {{{\left| {{\alpha _{{\rm{sd}},l,{\rho _{{\rm{sd}}}}}}} \right|}^2}} \right\}} \right)$ and ${\cal{CN}}\left( {0,N{\rm{E}}\left\{ {{{\left| {{\alpha _{{\rm{sr}},l{{,}}{\rho _{{\rm{sr}}}}}}} \right|}^2}} \right\} \cdot {\rm{E}}\left\{ {{{\left| {{\alpha _{{\rm{rd}},l,{\rho _{{\rm{rd}}}}}}} \right|}^2}} \right\}} \right)$, respectively, where ${\rm{E}}\left\{  \cdot  \right\}$ denotes the expectation operation. In the next section, the channel model and it corresponding distribution feature will be used for the performance analysis.

\section{BER Performance Analysis}
\label{sect:BER Performance Analysis}
In this section, the theoretical BER expressions of scheme I and scheme II in the proposed RIS-$M$-FM-DCSK system are derived over a multipath Rayleigh fading channel. Without loss of generality, we make two assumptions here. First, we assume that the channel gains are unequal\cite{2019Multi,9761198,9035610}, i.e., ${\rm{E}}\left\{ {{{\left| {{\alpha _{{\rm{sd}},l,1}}} \right|}^2}} \right\} \ne {\rm{E}}\left\{ {{{\left| {{\alpha _{{\rm{sd}},l,2}}} \right|}^2}} \right\} \ne  \cdots  \ne {\rm{E}}\left\{ {{{\left| {{\alpha _{{\rm{sd}},l,{\rho _{{\rm{sd}}}}}}} \right|}^2}} \right\}$, ${\rm{E}}\left\{ {{{\left| {{\alpha _{{\rm{sr}},l,1}}} \right|}^2}} \right\} \ne {\rm{E}}\left\{ {{{\left| {{\alpha _{{\rm{sr}},l,2}}} \right|}^2}} \right\} \ne  \cdots  \ne {\rm{E}}\left\{ {{{\left| {{\alpha _{{\rm{sr}},l,{\rho _{{\rm{sr}}}}}}} \right|}^2}} \right\}$, and ${\rm{E}}\left\{ {{{\left| {{\alpha _{{\rm{rd}},l,1}}} \right|}^2}} \right\} \ne {\rm{E}}\left\{ {{{\left| {{\alpha _{{\rm{rd}},l,2}}} \right|}^2}} \right\} \ne  \cdots  \ne {\rm{E}}\left\{ {{{\left| {{\alpha _{{\rm{rd}},l,{\rho _{{\rm{rd}}}}}}} \right|}^2}} \right\}$. Second, we assume that the symbol duration is much longer than the path delays, i.e., $0 \le {\tau _{{\rho _{{\rm{sd}}}}}}/{\tau _{{\rho _{{\rm{ca}}}}}} \ll SF$, thus the inter-symbol interference can be ignored\cite{9277910,9372875,9035610}.
\subsection{Scheme I} \label{sect:Scheme I in the RIS-$M$-FM-DCSK system}
The overall BER ${P_{\rm{I}}}$ is related to ${P_{{\rm{I,b}}}}$ and ${P_{{\rm{I,c}}}}$, where ${P_{{\rm{I,b}}}}$ and ${P_{{\rm{I,c}}}}$ are the BERs of the information bits $b$ and $\bf{c}$, respectively. Therefore, the overall BER ${P_{\rm{I}}}$ is given by
\begin{align}
{P_{\rm{I}}} = \frac{{{P_{{\rm{I,b}}}}}}{{{\Xi _{{\rm{I}},{\rm{tot}}}}}} + \frac{{{P_{{\rm{I,c}}}}n{{\log }_2}M}}{{{\Xi _{{\rm{I}},{\rm{tot}}}}}}.
\label{eq:15func}
\end{align}
Without loss of generality, we assume that the information bit $b$ equals $1$ in the following analysis.
\subsubsection{Derivation of ${P_{{\rm{I,b}}}}$} \label{sect:Derivation of pib}
For the derivation of ${P_{{\rm{I,b}}}}$, we further express Eq.~(\ref{eq:4func}) as Eq.~(\ref{eq:16func}).
\begin{figure*}[!tbp]
\begin{align}
{A_{\rm{d}}} &\approx {\rm{Real}}\left\{ {\sum\limits_{{i_1} = 1}^{n + 1} {\sum\limits_{{i_2} = 1}^\beta  {\left[ {D_{{\rm{sd}}}^{ - \varepsilon }\sum\limits_{{\rho _{{\rm{sd}}}} = 1}^{{L_{{\rm{sd}}}}} {{{\left| {{\Lambda _{{\rm{sd}},{\rho _{{\rm{sd}}}}}}} \right|}^2}{{\left| {{\vartheta _{{i_1} - 1}}} \right|}^2}{{\left| {{k_{{i_2}}}} \right|}^2} + \sqrt {D_{{\rm{sd}}}^{ - \varepsilon }} \sum\limits_{{\rho _{{\rm{sd}}}} = 1}^{{L_{{\rm{sd}}}}} {{\Lambda _{{\rm{sd}},{\rho _{{\rm{sd}}}}}}{\vartheta _{{i_1} - 1}}{k_{{i_2}}}w_{\left( {n + 1} \right)\beta  + \left( {{i_1} - 1} \right)\beta  + {i_2}}^*} } } \right.} } } \right\}\nonumber\\
 &= \sum\limits_{{i_1} = 1}^{n + 1} {\sum\limits_{{i_2} = 1}^\beta  {\left[ {D_{{\rm{sd}}}^{ - \varepsilon }\sum\limits_{{\rho _{{\rm{sd}}}} = 1}^{{L_{{\rm{sd}}}}} {{{\left| {{\Lambda _{{\rm{sd}},{\rho _{{\rm{sd}}}}}}} \right|}^2}{{\left| {{\vartheta _{{i_1} - 1}}} \right|}^2} \cdot {{\left| {{k_{{i_2}}}} \right|}^2}} } \right.} }  + \sqrt {D_{{\rm{sd}}}^{ - \varepsilon }} \sum\limits_{{\rho _{{\rm{sd}}}} = 1}^{{L_{{\rm{sd}}}}} {\left[ {\left( {{\Lambda _{{\rm{sd}},{\rho _{{\rm{sd}}}},1}}{k_{{i_2},1}}{\vartheta _{{i_1} - 1,1}} - {\Lambda _{{\rm{sd}},{\rho _{{\rm{sd}}}},1}}{k_{{i_2},2}}{\vartheta _{{i_1} - 1,2}}} \right.} \right.} \nonumber\\
&\left. { - {\Lambda _{{\rm{sd}},{\rho _{{\rm{sd}}}},2}}{k_{{i_2},1}}{\vartheta _{{i_1} - 1,2}} - {\Lambda _{{\rm{sd}},{\rho _{{\rm{sd}}}},2}}{k_{{i_2},2}}{\vartheta _{{i_1} - 1,1}}} \right){w_{\left( {n + 1} \right)\beta  + \left( {{i_1} - 1} \right)\beta  + {i_2},1}} + \left( {{\Lambda _{{\rm{sd}},{\rho _{{\rm{sd}}}},1}}{k_{{i_2},1}}{\vartheta _{{i_1} - 1,2}} + {\Lambda _{{\rm{sd}},{\rho _{{\rm{sd}}}},1}}{k_{{i_2},2}}{\vartheta _{{i_1} - 1,1}}} \right.\nonumber\\
&\left. {\left. { + \!{\Lambda _{{\rm{sd}},{\rho _{{\rm{sd}}}},2}}{k_{{i_2},1}}{\vartheta _{{i_1} \!- \!1,1}} \!- \!{\Lambda _{{\rm{sd}},{\rho _{{\rm{sd}}}},2}}{k_{{i_2},2}}{\vartheta _{{i_1} - 1,2}}}\! \right){w_{\left( {n\! + \!1} \right)\beta  + \left( {{i_1}\! -\! 1} \right)\beta  \!+\! {i_2},2}}} \right] \!\!+ \!\!\sqrt {\!D_{{\rm{sd}}}^{ - \varepsilon }}\! \sum\limits_{{\rho _{{\rm{sd}}}} = 1}^{{L_{{\rm{sd}}}}} \!\!{\left[ {\left( {{\Lambda _{{\rm{sd}},{\rho _{{\rm{sd}}}},1}}{k_{{i_2},1}}{\vartheta _{{i_1} \!-\! 1,1}}\!\! -\! \!{\Lambda _{{\rm{sd}},{\rho _{{\rm{sd}}}},1}}{k_{{i_2},2}}{\vartheta _{{i_1} \!-\! 1,2}}} \right.} \!\right.} \nonumber\\
&\left. { - {\Lambda _{{\rm{sd}},{\rho _{{\rm{sd}}}},2}}{k_{{i_2},1}}{\vartheta _{{i_1} - 1,2}} - {\Lambda _{{\rm{sd}},{\rho _{{\rm{sd}}}},2}}{k_{{i_2},2}}{\vartheta _{{i_1} - 1,1}}} \right){w_{\left( {{i_1} - 1} \right)\beta  + {i_2},1}} + \left( {{\Lambda _{{\rm{sd}},{\rho _{{\rm{sd}}}},1}}{k_{{i_2},1}}{\vartheta _{{i_1} - 1,2}} + {\Lambda _{{\rm{sd}},{\rho _{{\rm{sd}}}},1}}{k_{{i_2},2}}{\vartheta _{{i_1} - 1,1}}} \right.\nonumber\\
&\left. { + {\Lambda _{{\rm{sd}},{\rho _{{\rm{sd}}}},2}}{k_{{i_2},1}}{\vartheta _{{i_1} - 1,1}} - {\Lambda _{{\rm{sd}},{\rho _{{\rm{sd}}}},2}}{k_{{i_2},2}}{\vartheta _{{i_1} - 1,2}}} \right){w_{\left( {{i_1} - 1} \right)\beta  + {i_2},2}} + \left( {{w_{\left( {n + 1} \right)\beta  + \left( {{i_1} - 1} \right)\beta  + {i_2},1}}{w_{\left( {{i_1} - 1} \right)\beta  + {i_2},1}}} \right.\nonumber\\
&\left. {\left. { + {w_{\left( {n + 1} \right)\beta  + \left( {{i_1} - 1} \right)\beta  + {i_2},2}}{w_{\left( {{i_1} - 1} \right)\beta  + {i_2},2}}} \right)} \right].
\label{eq:16func}
\end{align}
\hrulefill
\end{figure*}
In Eq.~(\ref{eq:16func}), we define ${\Lambda _{{\rm{sd}},{\rho _{{\rm{sd}}}}}} = {\Lambda _{{\rm{sd}},{\rho _{{\rm{sd}}}},1}} + j{\Lambda _{{\rm{sd}},{\rho _{{\rm{sd}}}},2}}$, $\exp \left( {j{\varphi _{{i_1} - 1}}} \right) = {\vartheta _{{i_1} - 1}} = {\vartheta _{{i_1} - 1,1}} + j{\vartheta _{{i_1} - 1,2}}$, ${k_{{i_2}}} = {k_{{i_2},1}} + j{k_{{i_2},2}}$, and ${w_\Im } = {w_{\Im ,1}} + j{w_{\Im ,2}}$ ($1 \le \Im  \le 2\left( {n + 1} \right)\beta $), where $\vartheta _{{i_1} - 1,1}^2 + \vartheta _{{i_1} - 1,2}^2 = 1$, $\sum\nolimits_{{i_2} = 1}^\beta  {{{\left| {{k_{{i_2}}}} \right|}^2}}  = \frac{{{E_{\rm{s}}}}}{{2\left( {n + 1} \right)}}$ and ${E_{\rm{s}}}$ is the transmitted energy per symbol. More specifically, the average transmitted energy per bit is defined as ${E_{\rm{b}}} = \frac{{{E_{\rm{s}}}}}{{{\Xi _{{\rm{I}},{\rm{tot}}}}}}$ in scheme I. Referring to\cite{7332774,9761226,9035610}, for a large $\beta $, the random variable ${A_{\rm{d}}}$ follows a Gaussian distribution. Hence, the mean and variance of the decision metric ${A_{\rm{d}}}$ can be respectively calculated as
\begin{align}
{\rm{E}}\left\{ {{A_{\rm{d}}}} \right\} &= \left( {n + 1} \right) \cdot \frac{{D_{{\rm{sd}}}^{ - \varepsilon }\sum\limits_{{i_1}=1}^{n + 1} {\sum\limits_{{\rho _{{\rm{sd}}}} = 1}^{{L_{{\rm{sd}}}}} {{{\left| {{\Lambda _{{\rm{sd}},{\rho _{{\rm{sd}}}}}}} \right|}^2}{{\left| {{\vartheta _{{i_1} - 1}}} \right|}^2}{E_{\rm{s}}}} } }}{{2\left( {n + 1} \right)}}\nonumber\\
 &= \frac{{D_{{\rm{sd}}}^{ - \varepsilon }\sum\limits_{{\rho _{{\rm{sd}}}} = 1}^{{L_{{\rm{sd}}}}} {{{\left| {{\Lambda _{{\rm{sd}},{\rho _{{\rm{sd}}}}}}} \right|}^2}{E_{\rm{s}}}} }}{2},
 \label{eq:17func}
\end{align}
\begin{align}
{\rm{Var}}\left\{ {{A_{\rm{d}}}} \right\}& = \frac{{D_{{\rm{sd}}}^{ - \varepsilon }\sum\limits_{{i_1} = 1}^{n + 1} {\sum\limits_{{\rho _{{\rm{sd}}}} = 1}^{{L_{{\rm{sd}}}}} {{{\left| {{\Lambda _{{\rm{sd}},{\rho _{{\rm{sd}}}}}}} \right|}^2}{{\left| {{\vartheta _{{i_1} - 1}}} \right|}^2}{E_{\rm{s}}}{N_0}} } }}{{2\left( {n + 1} \right)}} \nonumber\\
&+ \frac{{\beta \sum\limits_{{i_1} = 1}^{n + 1} {N_0^2{{\left| {{\vartheta _{{i_1} - 1}}} \right|}^2}} }}{2}\nonumber\\
 &= \frac{{D_{{\rm{sd}}}^{ - \varepsilon }\sum\limits_{{\rho _{{\rm{sd}}}} = 1}^{{L_{{\rm{sd}}}}} {{{\left| {{\Lambda _{{\rm{sd}},{\rho _{{\rm{sd}}}}}}} \right|}^2}{E_{\rm{s}}}{N_0}}  + \beta N_0^2\left( {n + 1} \right)}}{2},
 \label{eq:18func}
\end{align}
where ${\rm{Var}}\left\{  \cdot  \right\}$ denotes the variance operation. In addition, we define ${r_{\rm{s}}} = \frac{{D_{{\rm{sd}}}^{ - \varepsilon }\sum\nolimits_{{\rho _{{\rm{sd}}}} = 1}^{{L_{{\rm{sd}}}}} {{{\left| {{\Lambda _{{\rm{sd}},{\rho _{{\rm{sd}}}}}}} \right|}^2}{E_{\rm{s}}}} }}{{{N_0}}}$ as the instantaneous received SNR for scheme I. Because ${A_{\rm{d}}}$ follows a Gaussian distribution with mean ${\rm{E}}\left\{ {{A_{\rm{d}}}} \right\}$ and variance ${\rm{Var}}\left\{ {{A_{\rm{d}}}} \right\}$, the BER of the information bit $b$ conditioned on ${r_{\rm{s}}}$ can be expressed as
\begin{align}
{P_{{\rm{I,b|}}{r_{\rm{s}}}}} &= \frac{1}{2}{\rm{erfc}}\left( {\frac{{{\rm{E}}\left\{ {{A_{\rm{d}}}} \right\}}}{{\sqrt {2{\rm{Var}}\left\{ {{A_{\rm{d}}}} \right\}} }}} \right)\nonumber\\
 &= \frac{1}{2}{\rm{erfc}}\left( {\frac{1}{{2\sqrt {r_s^{ - 1} + \beta \left( {n + 1} \right)r_s^{ - 2}} }}} \right),
 \label{eq:19func}
\end{align}
where ${\rm{erfc}}\left(  \cdot  \right)$ is the complementary error function\cite{2005Digital}. Next, the average BER of the information bit $b$ over a multipath Rayleigh fading channel is given by
\begin{align}
{P_{{\rm{I,b}}}} = \int_0^\infty  {{P_{{\rm{I,b|}}{r_{\rm{s}}}}}} f\left( {{r_{\rm{s}}}} \right)d{r_{\rm{s}}},
 \label{eq:20func}
\end{align}
where $f\left( {{r_{\rm{s}}}} \right)$ is the probability density function (PDF) of ${r_{\rm{s}}}$. The expression of $f\left( {{r_{\rm{s}}}} \right)$ is written as
\begin{align}
f\left( {{r_{\rm{s}}}} \right) = \sum\limits_{{\rho _{{\rm{sd}}}} = 1}^{L{ _{{\rm{sd}}}}} {\frac{{{\chi _{{\rho _{{\rm{sd}}}}}}}}{{{r_{{\rho _{{\rm{sd}}}},{\rm{ave}}}}}}} \exp \left( { - \frac{{{r_{\rm{s}}}}}{{{r_{{\rho _{{\rm{sd}}}},{\rm{ave}}}}}}} \right),
\label{eq:21func}
\end{align}
where
\begin{align}
{\chi _{{\rho _{{\rm{sd}}}}}} = \prod\limits_{\varsigma   = 1,\varsigma  \ne {\rho _{{\rm{sd}}}}}^{L{_{{\rm{sd}}}}} {\frac{{{r_{{\rho _{{\rm{sd}}}},{\rm{ave}}}}}}{{{r_{{\rho _{{\rm{sd}}}},{\rm{ave}}}} - {r_{\varsigma ,{\rm{ave}}}}}}} ,
\label{eq:22func}
\end{align}
in which ${r_{{\rho _{{\rm{sd}}}},{\rm{ave}}}} = E\left\{ {{{\left| {{\Lambda _{{\rm{sd}},{\rho _{{\rm{sd}}}}}}} \right|}^2}} \right\} \cdot \frac{{D_{{\rm{sd}}}^{ - \varepsilon }{E_{\rm{s}}}}}{{{N_0}}}$ is the average value of the instantaneous received SNR on the ${\rho _{{\rm{sd}}}}$-th path.
\subsubsection{Derivation of ${P_{{\rm{I,c}}}}$} \label{sect:Derivation of pic}
To derive ${P_{{\rm{I,c}}}}$, Eq.~(\ref{eq:6func}) is further expressed as
\begin{align}
{A_{{\rm{s}},\partial }} \approx {B_{{\rm{signal}}}}\exp \left( {j{\varphi _\partial }} \right) + {B_{{\rm{noise}}}},
\label{eq:23func}
\end{align}
where ${B_{{\rm{signal}}}}$ and ${B_{{\rm{noise}}}}$ are respectively written as
\begin{align}
{B_{{\rm{signal}}}} = 2D_{{\rm{sd}}}^{ - \varepsilon }\sum\limits_{{\rho _{{\rm{sd}}}} = 1}^{{L_{{\rm{sd}}}}} {{{\left| {{\Lambda _{{\rm{sd}},{\rho _{{\rm{sd}}}}}}} \right|}^2}} {\left| {{k_i}} \right|^2},
\label{eq:24func}
\end{align}
\begin{align}
&{B_{{\rm{noise}}}} \nonumber\\
&= \sum\limits_{i = 1}^\beta \!\! {\left[\! {\left( {\sqrt {D_{{\rm{sd}}}^{ - \varepsilon }} \sum\limits_{{\rho _{{\rm{sd}}}} = 1}^{{L_{{\rm{sd}}}}} {\Lambda _{{\rm{sd}},{\rho _{{\rm{sd}}}}}^*k_i^ * } } \right)\! \times\! } \right.} \left( {{w_{\partial \beta \! + \!i}} + {w_{\left( {n + 1} \right)\beta \! +\! \partial \beta  + i}}} \right)\nonumber\\
 &+ \left( {\sqrt {D_{{\rm{sd}}}^{ - \varepsilon }} \sum\limits_{{\rho _{{\rm{sd}}}} = 1}^{{L_{{\rm{sd}}}}} {{\Lambda _{{\rm{sd}},{\rho _{{\rm{sd}}}}}}\exp \left( {j{\varphi _\partial }} \right){k_i}} } \right) \!\times\! \left( {w_i^* + w_{\left( {n + 1} \right)\beta + i}^*} \right)\nonumber\\
&\left. { + \left( {w_i^*{w_{\partial \beta  + i}} + w_{\left( {n + 1} \right)\beta  + i}^*{w_{\left( {n + 1} \right)\beta  + \partial \beta  + i}}} \right)} \right].
\label{eq:25func}
\end{align}
For the demodulation of the $M$-PSK symbol $\exp \left( {j{\varphi _\partial }} \right)$, ${B_{{\rm{signal}}}}$, and ${B_{{\rm{noise}}}}$ can be regarded as the desired component and noise component, respectively. According to\cite{John1983Digital}, the equivalent conditional received SNR of the $M$-PSK symbol $\exp \left( {j{\varphi _\partial }} \right)$ is given by
\begin{align}
{r_{{\rm{s}},{\rm{equ}}{\rm{|}}{r_{\rm{s}}}}} = \frac{{{{\rm{E}}^2}\left\{ {{B_{{\rm{signal}}}}} \right\}}}{{{\rm{Var}}\left\{ {{B_{{\rm{noise}}}}} \right\}}},
\label{eq:26func}
\end{align}
where ${\rm{E}}\left\{ {{B_{{\rm{signal}}}}} \right\}$ and ${\rm{Var}}\left\{ {{B_{{\rm{noise}}}}} \right\}$ can be respectively calculated as
\begin{align}
{\rm{E}}\left\{ {{B_{{\rm{signal}}}}} \right\} = \frac{{D_{{\rm{sd}}}^{ - \varepsilon }\sum\limits_{{\rho _{{\rm{sd}}}} = 1}^{{L_{{\rm{sd}}}}} {{{\left| {{\Lambda _{{\rm{sd}},{\rho _{{\rm{sd}}}}}}} \right|}^2}} {E_{\rm{s}}}}}{{n + 1}},
\label{eq:27func}
\end{align}
\begin{align}
{\rm{Var}}\left\{ {{B_{{\rm{noise}}}}} \right\} = \frac{{2D_{{\rm{sd}}}^{ - \varepsilon }\sum\limits_{{\rho _{{\rm{sd}}}} = 1}^{{L_{{\rm{sd}}}}} {{{\left| {{\Lambda _{{\rm{sd}},{\rho _{{\rm{sd}}}}}}} \right|}^2}} {E_{\rm{s}}}{N_0}}}{{n + 1}} + 2\beta N_0^2.
\label{eq:28func}
\end{align}
Substituting Eqs.~(\ref{eq:27func}) and (\ref{eq:28func}) into Eq.~(\ref{eq:26func}), one can obtain
\begin{align}
{r_{{\rm{s}},{\rm{equ}}{\rm{|}}{r_{\rm{s}}}}} = \frac{1}{{2\left( {n + 1} \right)r_{\rm{s}}^{ - 1} + 2\beta {{\left( {n + 1} \right)}^2}r_{\rm{s}}^{ - 2}}}.
\label{eq:29func}
\end{align}
Afterwards, according to\cite{John1983Digital}, the conditional BER of the information bits $\bf c$ is given by
\begin{align}
{P_{{\rm{I,c|}}{r_{\rm{s}}}}} = \left\{ \begin{array}{l}
Q\left( {\sqrt {2{r_{{\rm{s}},{\rm{equ}}{\rm{|}}{r_{\rm{s}}}}}} } \right),{\rm{                                   }}\ \ \ \ \ \ \ \ \ \ \ \ \ \ \ \ \ \ \ M = 2\\
\frac{2}{{{{\log }_2}M}}Q\left( {\sqrt {2{{\sin }^2}\left( {\frac{\pi }{M}} \right){r_{{\rm{s}},{\rm{equ}}{\rm{|}}{r_{\rm{s}}}}}} } \right),{\rm{     }}M \ge 4
\end{array} \right.,
\label{eq:30func}
\end{align}
where $Q\left(  \cdot  \right)$ is the Gaussian $Q$-function\cite{2005Digital}. Next, combining Eqs.~(\ref{eq:21func}) and (\ref{eq:30func}), the average BER of the information bits $\bf c$ over a multipath Rayleigh fading channel is given by
\begin{align}
{P_{{\rm{I,c}}}} = \int_0^\infty  {{P_{{\rm{I,c|}}{r_{\rm{s}}}}}} f\left( {{r_{\rm{s}}}} \right)d{r_{\rm{s}}}.
\label{eq:31func}
\end{align}
Finally, substituting Eqs.~(\ref{eq:20func}) and (\ref{eq:31func}) into Eq.~(\ref{eq:15func}), one can obtain the overall BER of scheme I over a multipath Rayleigh fading channel.
\subsection{Scheme II} \label{sect:Scheme II in the RIS-M-FM-DCSK system}
In scheme II, one information bit $\dot b$ and $n{\log _2}M$ information bits $\dot {\bf{c}}$ are sent from the source and relay to the destination, respectively, in each transmission period. Here, we define the BERs of the information bits $\dot b$ and $\dot {\bf{c}}$ as ${P_{{\rm{II,\dot b}}}}$ and ${P_{{\rm{II,\dot c}}}}$, respectively. Without loss of generality, we assume that the information bit $\dot b$ equals $1$ in the following analysis.
\subsubsection{Derivation of ${P_{{\rm{II,\dot b}}}}$} \label{sect:Derivation of dotpib}
To derive ${P_{{\rm{II,\dot b}}}}$, Eq.~(\ref{eq:12func}) is further expressed as Eq.~(\ref{eq:32func}).
\begin{figure*}[!tbp]
\begin{align}
{{\dot A}_{\rm{d}}} &\approx {\rm{Real}}\left\{ {\sum\limits_{{i_1} = 1}^{n + 1} {\sum\limits_{{i_2} = 1}^\beta  {\left[ {D_{{\rm{sr}}}^{ - \varepsilon }D_{{\rm{rd}}}^{ - \varepsilon }\sum\limits_{{\rho _{{\rm{ca}}}} = 1}^{{L_{{\rm{tot}}}}} {{{\left| {{\Gamma _{{\rm{sd,}}{\rho _{{\rm{ca}}}}}}} \right|}^2}} } \right.} } } \right.{\left| {{{\dot \vartheta }_{{i_1} - 1}}} \right|^2}{\left| {{k_{{i_2}}}} \right|^2} + \sqrt {D_{{\rm{sr}}}^{ - \varepsilon }D_{{\rm{rd}}}^{ - \varepsilon }} \sum\limits_{{\rho _{{\rm{ca}}}} = 1}^{{L_{{\rm{tot}}}}} {{\Gamma _{{\rm{sd,}}{\rho _{{\rm{ca}}}}}}{{\dot \vartheta }_{{i_1} - 1}}{k_{{i_2}}}w_{\left( {n + 1} \right)\beta  + \left( {{i_1} - 1} \right)\beta  + {i_2}}^*} \nonumber\\
&\left. {\left. { + \sqrt {D_{{\rm{sr}}}^{ - \varepsilon }D_{{\rm{rd}}}^{ - \varepsilon }} \sum\limits_{{\rho _{{\rm{ca}}}} = 1}^{{L_{{\rm{sd}}}}} {\Gamma _{{\rm{sd,}}{\rho _{{\rm{ca}}}}}^*\dot \vartheta _{{i_1} - 1}^*k_{{i_2}}^*{w_{\left( {{i_1} - 1} \right)\beta  + {i_2}}} + } {w_{\left( {{i_1} - 1} \right)\beta  + {i_2}}}w_{\left( {n + 1} \right)\beta  + \left( {{i_1} - 1} \right)\beta  + {i_2}}^*} \right]} \right\}\nonumber\\
& = \sum\limits_{{i_1} = 1}^{n + 1} {\sum\limits_{{i_2} = 1}^\beta  {\left[ {D_{{\rm{sr}}}^{ - \varepsilon }D_{{\rm{rd}}}^{ - \varepsilon }\sum\limits_{{\rho _{{\rm{ca}}}} = 1}^{{L_{{\rm{tot}}}}} {{{\left| {{\Gamma _{{\rm{sd,}}{\rho _{{\rm{ca}}}}}}} \right|}^2}{{\left| {{{\dot \vartheta }_{{i_1} - 1}}} \right|}^2} \cdot {{\left| {{k_{{i_2}}}} \right|}^2}} } \right.} }  + \sqrt {D_{{\rm{sr}}}^{ - \varepsilon }D_{{\rm{rd}}}^{ - \varepsilon }} \sum\limits_{{\rho _{{\rm{ca}}}} = 1}^{{L_{{\rm{tot}}}}} {\left[ {\left( {{\Gamma _{{\rm{sd,}}{\rho _{{\rm{ca}}}},1}}{k_{{i_2},1}}{{\dot \vartheta }_{{i_1} - 1,1}} - {\Gamma _{{\rm{sd,}}{\rho _{{\rm{ca}}}},1}}{k_{{i_2},2}}{{\dot \vartheta }_{{i_1} - 1,2}}} \right.} \right.} \nonumber\\
&\left. { - \!{\Gamma _{{\rm{sd}},{\rho\! _{{\rm{ca}}}},2}}{k_{{i_2},\!1}}{{\dot \vartheta }_{{i_1} \!-\! 1\!,2}}\!\! -\!\! {\Gamma \!_{{\rm{sd}},{\rho _{{\rm{ca}}}},2}}{k_{{i_2},2}}{{\dot \vartheta }_{{i_1} - 1,1}}} \!\right)\!\!{w_{\left( {n\! +\! 1} \!\right)\!\beta \! +\! \left( {{i_1} \!-\! 1} \right)\beta \! + \!{i_2},1}} \!\!+\!\! \left(\! {{\Gamma \!_{{\rm{sd}},{\rho _{{\rm{ca}}}},1}}\!{k_{{i_2},1}}{{\dot \vartheta }_{{i_1} \!-\! 1,2}} \!\!+ \!\!{\Gamma \!_{{\rm{sd}},{\rho _{{\rm{ca}}}},1}}\!{k_{{i_2},2}}{{\dot \vartheta }_{{i_1} \!-\! 1,1}} \!\!+\!\! {\Gamma \!_{{\rm{sd}},{\rho _{{\rm{ca}}}},2}}{k_{{i_2},1}}{{\dot \vartheta }_{{i_1}\! - \!1,1}}} \right.\nonumber\\
&\left. {\left. { - {\Gamma _{{\rm{sd}},{\rho _{{\rm{ca}}}},2}}{k_{{i_2},2}}{{\dot \vartheta }_{{i_1} \!-\! 1,2}}} \right){w_{\left( {n + 1} \right)\beta  + \left( {{i_1} - 1} \right)\beta  + {i_2},2}}} \right] + \sqrt {D_{{\rm{sr}}}^{ - \varepsilon }D_{{\rm{rd}}}^{ - \varepsilon }} \sum\limits_{{\rho _{{\rm{sd}}}} = 1}^{{L_{{\rm{tot}}}}} {\left[ {\left( {{\Gamma _{{\rm{sd,}}{\rho _{{\rm{ca}}}},1}}{k_{{i_2},1}}{{\dot \vartheta }_{{i_1} - 1,1}} - {\Gamma _{{\rm{sd,}}{\rho _{{\rm{ca}}}},1}}{k_{{i_2},2}}{{\dot \vartheta }_{{i_1} - 1,2}}} \right.} \right.} \nonumber\\
&\left. { -\! {\Gamma _{{\rm{sd}},{\rho _{{\rm{ca}}}},2}}{k_{{i_2},1}}{{\dot \vartheta }_{{i_1} \!- \!1,2}} \!- \!{\Gamma _{{\rm{sd,}}{\rho _{{\rm{ca}}}},2}}{k_{{i_2},2}}{{\dot \vartheta }_{{i_1} \!-\! 1,1}}} \right)\!{w_{\left( {{i_1}\! -\! 1} \right)\beta \! +\! {i_2},1}}\! + \!\left(\! {{\Gamma _{{\rm{sd,}}{\rho _{{\rm{ca}}}},1}}{k_{{i_2},1}}{{\dot \vartheta }_{{i_1} \!-\! 1,2}} \!+\! {\Gamma\! _{{\rm{sd}},{\rho _{{\rm{ca}}}},1}}{k_{{i_2},2}}{{\dot \vartheta }_{{i_1} \!-\! 1,1}}} \right. \!+\! {\Gamma \!_{{\rm{sd}},{\rho _{{\rm{ca}}}},2}}{k_{{i_2},1}}{{\dot \vartheta }_{{i_1} \!-\! 1,1}}\nonumber\\
&\left. {\left. {\left. {\! - {\Gamma _{{\rm{sd,}}{\rho _{{\rm{ca}}}},2}}{k_{{i_2},2}}{{\dot \vartheta }_{{i_1} - 1,2}}} \right){w_{\left( {{i_1}\! - \!1} \right)\beta  \!+\! {i_2},2}}} \right] \!+\! \left( {{w_{\left( {n + 1} \right)\beta \! + \!\left( {{i_1} - 1} \right)\beta  + {i_2},1}}{w_{\left( {{i_1} - 1} \right)\beta  + {i_2},1}} + {w_{\left( {n + 1} \right)\beta  \!+\! \left( {{i_1} - 1} \right)\beta \! + \!{i_2},2}}{w_{\left( {{i_1} - 1} \right)\beta  \!+\! {i_2},2}}} \right)} \right].
\label{eq:32func}
\end{align}
\hrulefill
\end{figure*}
In Eq.~(\ref{eq:32func}), we define ${\Gamma _{{\rm{sd}},{\rho _{{\rm{sd}}}}}} = {\Gamma _{{\rm{sd,}}{\rho _{{\rm{ca}}}},1}} + j{\Gamma _{{\rm{sd,}}{\rho _{{\rm{ca}}}},2}}$, $\exp \left( {j{{\dot \varphi }_{{i_1} - 1}}} \right) = {\dot \vartheta _{{i_1} - 1}} = {\dot \vartheta _{{i_1} - 1,1}} + j{\dot \vartheta _{{i_1} - 1,2}}$, where $\dot \vartheta _{{i_1} - 1,1}^2 + \dot \vartheta _{{i_1} - 1,2}^2 = 1$. Similar to the variable ${A_{\rm{d}}}$, the decision metric ${\dot A_{\rm{d}}}$ follows the Gaussian distribution. Hence, the mean and variance are respectively calculated as
\begin{align}
{\rm{E}}\left\{ {{{\dot A}_{\rm{d}}}} \right\} &= \left( {n + 1} \right) \cdot \frac{{D_{{\rm{sr}}}^{ - \varepsilon }D_{{\rm{rd}}}^{ - \varepsilon }\sum\limits_{{i_1}}^{n + 1} {\sum\limits_{{\rho _{{\rm{ca}}}} = 1}^{{L_{{\rm{tot}}}}} {{{\left| {{\Gamma _{{\rm{sd,}}{\rho _{{\rm{ca}}}}}}} \right|}^2}{{\left| {{{\dot \vartheta }_{{i_1} - 1}}} \right|}^2}{E_{\rm{s}}}} } }}{{2\left( {n + 1} \right)}}\nonumber\\
 &= \frac{{D_{{\rm{sr}}}^{ - \varepsilon }D_{{\rm{rd}}}^{ - \varepsilon }\sum\limits_{{\rho _{{\rm{ca}}}} = 1}^{{L_{{\rm{tot}}}}} {{{\left| {{\Gamma _{{\rm{sd,}}{\rho _{{\rm{ca}}}}}}} \right|}^2}{E_{\rm{s}}}} }}{2},
 \label{eq:33func}
\end{align}
\begin{align}
{\rm{Var}}\left\{ {{{\dot A}_{\rm{d}}}} \right\} &= \frac{{D_{{\rm{sr}}}^{ - \varepsilon }D_{{\rm{rd}}}^{ - \varepsilon }\sum\limits_{{i_1} = 1}^{n + 1} {\sum\limits_{{\rho _{{\rm{ca}}}} = 1}^{{L_{{\rm{tot}}}}} {{{\left| {{\Gamma _{{\rm{sd,}}{\rho _{{\rm{ca}}}}}}} \right|}^2}{{\left| {{{\dot \vartheta }_{{i_1} - 1}}} \right|}^2}{E_{\rm{s}}}{N_0}} } }}{{2\left( {n + 1} \right)}} \nonumber\\
&+ \frac{{\beta \sum\limits_{{i_1} = 1}^{n + 1} {N_0^2{{\left| {{{\dot \vartheta }_{{i_1} - 1}}} \right|}^2}} }}{2}\nonumber\\
 &= \frac{{D_{{\rm{sr}}}^{ - \varepsilon }D_{{\rm{rd}}}^{ - \varepsilon }\sum\limits_{{\rho _{{\rm{ca}}}} = 1}^{{L_{{\rm{tot}}}}} {{{\left| {{\Gamma _{{\rm{sd,}}{\rho _{{\rm{ca}}}}}}} \right|}^2}{E_{\rm{s}}}{N_0}}  + \beta N_0^2\left( {n + 1} \right)}}{2}.
 \label{eq:34func}
\end{align}
Also, we define ${\dot r_{\rm{s}}} = \frac{{D_{{\rm{sr}}}^{ - \varepsilon }D_{{\rm{rd}}}^{ - \varepsilon }\sum\limits_{{\rho _{{\rm{ca}}}} = 1}^{{L_{{\rm{tot}}}}} {{{\left| {{\Gamma _{{\rm{sd,}}{\rho _{{\rm{ca}}}}}}} \right|}^2}{E_{\rm{s}}}} }}{{{N_0}}}$ as the instantaneous received SNR for scheme II. In addition, the relationship between the transmitted energy per symbol ${E_{\rm{s}}}$ and average transmitted energy per bit ${E_{\rm{b}}}$ is ${E_{\rm{s}}} = \left( {{\Xi _{{\rm{II}},{\rm{s}}}} + {\Xi _{{\rm{II}},{\rm{r}}}}} \right){E_{\rm{b}}}$ in scheme II. Since the random variable ${\dot A_{\rm{d}}}$ follows the Gaussian distribution with mean ${\rm{E}}\left\{ {{{\dot A}_{\rm{d}}}} \right\}$ and variance ${\rm{Var}}\left\{ {{{\dot A}_{\rm{d}}}} \right\}$, the conditional BER of the information bit $\dot b$ can be expressed as
\begin{align}
{P_{{\rm{II,\dot b|}}{{\dot r}_{\rm{s}}}}} &= \frac{1}{2}{\rm{erfc}}\left( {\frac{{{\rm{E}}\left\{ {{{\dot A}_{\rm{d}}}} \right\}}}{{\sqrt {2{\rm{Var}}\left\{ {{{\dot A}_{\rm{d}}}} \right\}} }}} \right)\nonumber\\
 &= \frac{1}{2}{\rm{erfc}}\left( {\frac{1}{{2\sqrt {\dot r_{\rm{s}}^{ - 1} + \beta \left( {n + 1} \right)\dot r_{\rm{s}}^{ - 2}} }}} \right).
 \label{eq:35func}
\end{align}
Because ${\dot r_{\rm{s}}}$ and ${r_{\rm{s}}}$ follow the same distribution, the PDF of ${\dot r_{\rm{s}}}$ can be attained by employing Eq.~(\ref{eq:21func}). Next, the average BER for the transmitted bits of the source over a multipath Rayleigh fading channel is given by
\begin{align}
{P_{{\rm{II,\dot b}}}} = \int_0^\infty  {{P_{{\rm{II,\dot b|}}{{\dot r}_{\rm{s}}}}}} f\left( {{{\dot r}_{\rm{s}}}} \right)d{\dot r_{\rm{s}}}.
 \label{eq:36func}
\end{align}
\subsubsection{Derivation of ${P_{{\rm{II,\dot c}}}}$} \label{sect:Derivation of dotpiic}
To derive the BER of the transmitted bits from the relay to the destination ${P_{{\rm{II,\dot c}}}}$, Eq.~(\ref{eq:14func}) is further expressed as
\begin{align}
{\dot A_{{\rm{s}},\partial }} \approx {\dot B_{{\rm{signal}}}}\exp \left( {j{{\dot \varphi }_\partial }} \right) + {\dot B_{{\rm{noise}}}},
 \label{eq:37func}
\end{align}
where ${\dot B_{{\rm{signal}}}}$ and ${\dot B_{{\rm{noise}}}}$ are respectively calculated as
\begin{align}
{\dot B_{{\rm{signal}}}} = 2D_{{\rm{sr}}}^{ - \varepsilon }D_{{\rm{rd}}}^{ - \varepsilon }\sum\limits_{{\rho _{{\rm{ca}}}} = 1}^{{L_{{\rm{tot}}}}} {{{\left| {{\Gamma _{{\rm{sd,}}{\rho _{{\rm{ca}}}}}}} \right|}^2}} {\left| {{k_i}} \right|^2},
 \label{eq:38func}
\end{align}
\begin{align}
&{{\dot B}_{{\rm{noise}}}}\nonumber\\
&= \!\sum\limits_{i = 1}^\beta \! {\left[\! {\left( \!{\sqrt {D_{{\rm{sr}}}^{ - \varepsilon }D_{{\rm{rd}}}^{ - \varepsilon }} \sum\limits_{{\rho _{{\rm{ca}}}} = 1}^{{L_{{\rm{tot}}}}} {\Gamma _{{\rm{sd,}}{\rho _{{\rm{ca}}}}}^*k_i^ * } } \right)} \right.} \!\!\! \times\!\! \left( {{w_{\partial \beta \! +\! i}}\! + \!{w_{\left( {n \!+ \!1} \right)\beta  \!+\! \partial \beta \! + \!i}}} \right)\nonumber\\
 &+ \!\sum\limits_{i = 1}^\beta \!\! {\left(\! \!{\sqrt {\!D_{{\rm{sr}}}^{ - \varepsilon }\!D_{{\rm{rd}}}^{ - \varepsilon }}\! \sum\limits_{{\rho _{{\rm{ca}}}} = 1}^{{L_{{\rm{tot}}}}}\!\! {{\Gamma \!_{{\rm{sd}},{\rho _{{\rm{ca}}}}}}\!\exp \left( {j{{\dot \varphi }_\partial }} \right){k_i}} } \right)} \!\!\! \times\!\! \left( {w_i^*\! +\! w_{\left( {n \!+\! 1} \right)\beta\!  +\! i}^*} \right)\nonumber\\
&\left. { + \sum\limits_{i = 1}^\beta  {\left( {w_i^*{w_{\partial \beta  + i}} + w_{\left( {n + 1} \right)\beta  + i}^*{w_{\left( {n + 1} \right)\beta  + \partial \beta  + i}}} \right)} } \right].
 \label{eq:39func}
\end{align}
The signals ${\dot B_{{\rm{signal}}}}$ and ${\dot B_{{\rm{noise}}}}$ are the desired component and noise component, respectively, for the demodulation of the $M$-PSK symbol $\exp \left( {j{{\dot \varphi }_\partial }} \right)$. Hence, the equivalent conditional received SNR of the $M$-PSK symbol $\exp \left( {j{{\dot \varphi }_\partial }} \right)$ can be formulated as\cite{John1983Digital}
\begin{align}
{\dot r_{{\rm{s}},{\rm{equ}}{\rm{|}}{{\dot r}_{\rm{s}}}}} = \frac{{{{\rm{E}}^2}\left\{ {{{\dot B}_{{\rm{signal}}}}} \right\}}}{{{\rm{Var}}\left\{ {{{\dot B}_{{\rm{noise}}}}} \right\}}},
 \label{eq:40func}
\end{align}
where ${\rm{E}}\left\{ {{{\dot B}_{{\rm{signal}}}}} \right\}$ and ${\rm{Var}}\left\{ {{{\dot B}_{{\rm{noise}}}}} \right\}$ are respectively
\begin{align}
{\rm{E}}\left\{ {{{\dot B}_{{\rm{signal}}}}} \right\} = \frac{{D_{{\rm{sr}}}^{ - \varepsilon }D_{{\rm{rd}}}^{ - \varepsilon }\sum\limits_{{\rho _{{\rm{ca}}}} = 1}^{{L_{{\rm{tot}}}}} {{{\left| {{\Gamma _{{\rm{sd,}}{\rho _{{\rm{ca}}}}}}} \right|}^2}} {E_{\rm{s}}}}}{{n + 1}},
 \label{eq:41func}
\end{align}
\begin{align}
{\rm{Var}}\left\{ {{{\dot B}_{{\rm{noise}}}}} \right\} = \frac{{2D_{{\rm{sr}}}^{ - \varepsilon }D_{{\rm{rd}}}^{ - \varepsilon }\sum\limits_{{\rho _{{\rm{ca}}}} = 1}^{{L_{{\rm{tot}}}}} {{{\left| {{\Gamma _{{\rm{sd,}}{\rho _{{\rm{ca}}}}}}} \right|}^2}} {E_{\rm{s}}}{N_0}}}{{n + 1}} + 2\beta N_0^2.
 \label{eq:42func}
\end{align}
Substituting Eqs.~(\ref{eq:41func}) and (\ref{eq:42func}) into Eq.~(\ref{eq:40func}), one can obtain
\begin{align}
{\dot r_{{\rm{s}},{\rm{equ}}{\rm{|}}{{\dot r}_{\rm{s}}}}} = \frac{1}{{2\left( {n + 1} \right)\dot r_{\rm{s}}^{ - 1} + 2\beta {{\left( {n + 1} \right)}^2}\dot r_{\rm{s}}^{ - 2}}}.
 \label{eq:43func}
\end{align}
Then, according to\cite{John1983Digital}, the conditional BER of the information bits $\dot {\bf{c}}$ is given by
\begin{align}
{P_{{\rm{II,\dot c|}}{{\dot r}_{\rm{s}}}}} = \left\{ \begin{array}{l}
Q\left( {\sqrt {2{{\dot r}_{{\rm{s}},{\rm{equ}}{\rm{|}}{{\dot r}_{\rm{s}}}}}} } \right),\ \ \ \ \ \ \ \ \ \ \ \ \ \ \ \ \ \ \ M = 2\\
\frac{2}{{{{\log }_2}M}}Q\left( {\sqrt {2{{\sin }^2}\left( {\frac{\pi }{M}} \right){{\dot r}_{{\rm{s}},{\rm{equ}}{\rm{|}}{{\dot r}_{\rm{s}}}}}} } \right),{\rm{  }}M \ge 4
\end{array} \right..
 \label{eq:44func}
\end{align}
Finally, the average BER for the transmitted bits for the relay over a multipath Rayleigh fading channel is formulated as
\begin{align}
{P_{{\rm{II,\dot c}}}} = \int_0^\infty  {{P_{{\rm{II,\dot c|}}{{\dot r}_{\rm{s}}}}}} f\left( {{{\dot r}_{\rm{s}}}} \right)d{\dot r_{\rm{s}}}.
 \label{eq:45func}
\end{align}

\section{Simulation Results and Discussion}\label{sect:Simulation Results and Discussion}
In this section, we evaluate the performance of the proposed RIS-$M$-FM-DCSK system with scheme I and scheme II by simulations. In simulations, we consider the following parameter setting for the multipath fading channels. In the CS-i, for the link RIS$ \to $destination, the parameters of the multipath Rayleigh fading channel are set to ${L_{{\rm{sd}}}} = 2$, ${\rm{E}}\left\{ {{{\left| {{\alpha _{{\rm{sd}},l,1}}} \right|}^2}} \right\} = \frac{2}{3}$, ${\rm{E}}\left\{ {{{\left| {{\alpha _{{\rm{sd}},l,2}}} \right|}^2}} \right\} = \frac{1}{3}$,
${\tau _1} = 0$, and ${\tau _2} = 3{T_{\rm{c}}}$. In the CS-ii, for the link source$ \to $relay, the parameters of the multipath Rayleigh fading channel are set to ${L_{{\rm{sr}}}} = 3$, ${\rm{E}}\left\{ {{{\left| {{\alpha _{{\rm{sr}},l,1}}} \right|}^2}} \right\} = \frac{1}{2}$, ${\rm{E}}\left\{ {{{\left| {{\alpha _{{\rm{sr}},l,2}}} \right|}^2}} \right\} = \frac{1}{3}$, ${\rm{E}}\left\{ {{{\left| {{\alpha _{{\rm{sr}},l,3}}} \right|}^2}} \right\} = \frac{1}{6}$, ${\tau _1} = 0$, ${\tau _2} = 2{T_{\rm{c}}}$, and ${\tau _3} = 4{T_{\rm{c}}}$; for the link relay$ \to $destination, the parameters of the multipath Rayleigh fading channel are set to ${L_{{\rm{rd}}}} = 2$, ${\rm{E}}\left\{ {{{\left| {{\alpha _{{\rm{rd}},l,1}}} \right|}^2}} \right\} = \frac{4}{7}$, ${\rm{E}}\left\{ {{{\left| {{\alpha _{{\rm{rd}},l,2}}} \right|}^2}} \right\} = \frac{3}{7}$, ${\tau _1} = 0$, and ${\tau _2} = {T_{\rm{c}}}$. Furthermore, the path loss coefficient $\varepsilon $ is set to $2 $ in both CSs\cite{John1983Digital}.
\subsection{Performance Evaluation of the Proposed RIS-$M$-FM-DCSK system} \label{sect:Performance Evaluation of the Proposed RIS-$M$-FM-DCSK system}
\subsubsection{Scheme I} \label{sect:Scheme I}
\begin{figure}[tbp]
\center
\includegraphics[width=3.2in,height=2.56in]{{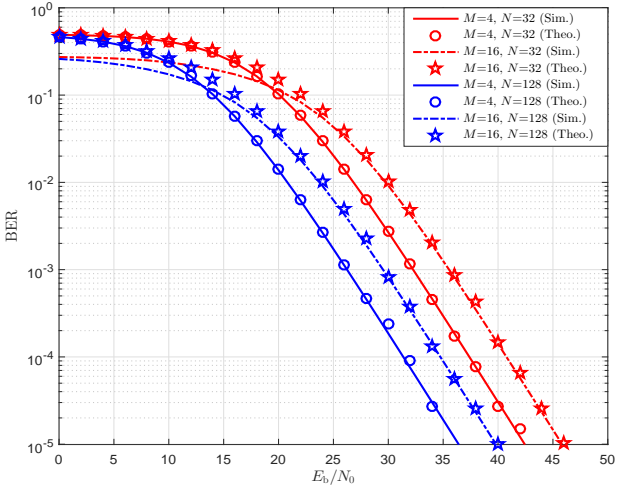}}
\vspace{-0.2cm}
\caption{Theoretical and simulated BER performance of scheme I in the RIS-$M$-FM-DCSK system over a multipath Rayleigh fading channel, where $SF = 300$, $n = 2$, and ${D_{{\rm{sd}}}} = 16{\rm{m}}$.}
\label{fig:fig2}  
\vspace{-2mm}
\end{figure}
\begin{figure}[tbp]
\center
\includegraphics[width=3.2in,height=2.56in]{{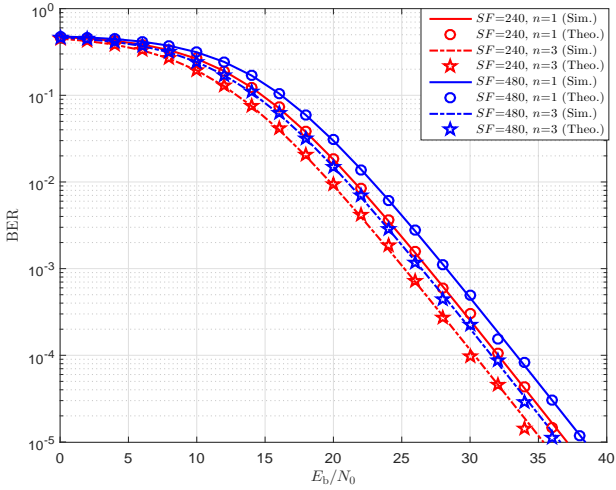}}
\vspace{-0.2cm}
\caption{Theoretical and simulated BER performance of scheme I in the RIS-$M$-FM-DCSK system over a multipath Rayleigh fading channel, where $SF = 300$, $M = 4$, $N = 128$, and ${D_{{\rm{sd}}}} = 16{\rm{m}}$.}
\label{fig:fig3}  
\end{figure}
Fig.~\ref{fig:fig2} shows the BER results of scheme I with different values of $M$ and $N$ over a multipath Rayleigh channel, where $n$ is set to 2, $SF = 300$, and ${D_{{\rm{sd}}}} = 16{\rm{m}}$. It can be seen that the theoretical BER results agree very well with the simulated ones in the BER region of practical interest. As shown in Fig.~\ref{fig:fig2}, the BER performance of scheme I is deteriorated with an increasing $M$. For instance, when $N = 128$, scheme I with $M = 4$ has about $2.5~{\rm{dB}}$ performance gain over scheme I with $M = 16$ at a BER of ${10^{ - 4}}$ over a multipath Rayleigh channel. Moreover, as expected, we can observe that the BER performance of scheme I improves as $N$ increases.  Specifically, when $M = 16$, the BER performance gap between $N = 32$ and $N = 128$ is about $5~{\rm{dB}}$ at a BER of ${10^{ - 4}}$.

Fig.~\ref{fig:fig3} shows BER performance of scheme I in the proposed RIS-$M$-FM-DCSK system with different $n$ and $SF$ over a multipath Rayleigh channel. It can be seen that the simulated BER results match well with the theoretical BER results. From this figure, one can observe that the BER of scheme I as $n$ increases. The main reason is that, when $n$ increases, less energy of the reference-part signal is required to transmit an information bit. More specifically, the transmitted energy of the reference part within the signal is shared by $1 + n{\log _2}M$ bits. Besides, the BER of scheme I becomes worse with the increase of $SF$, due to the fact that severer noise appears in the detection process as $SF$ increases.

\begin{figure}[tbp]
\center
\subfigure[\hspace{-0.8cm}]{ \label{fig:subfig:4a}
\includegraphics[width=3.2in,height=2.56in]{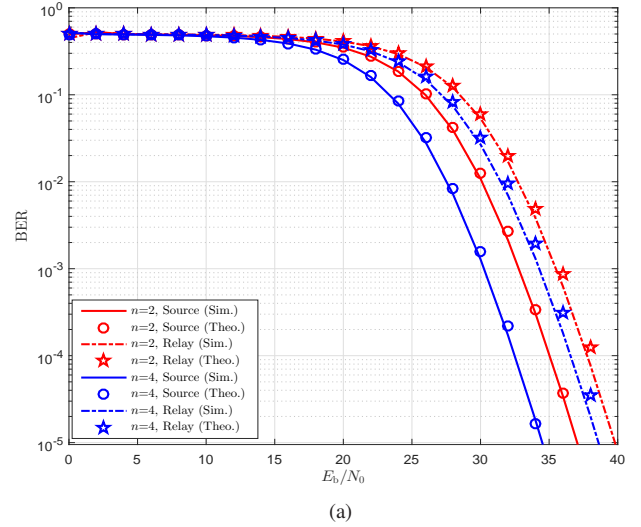}}
\subfigure[\hspace{-0.8cm}]{ \label{fig:subfig:4b}
\includegraphics[width=3.2in,height=2.56in]{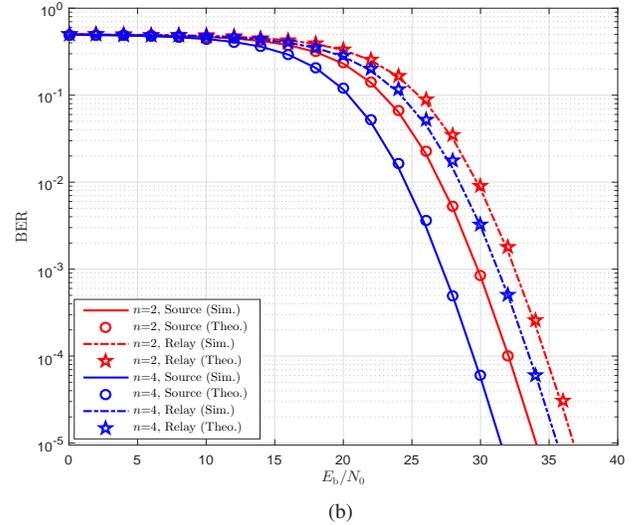}}
\caption{Theoretical and simulated BER performance of scheme II in the RIS-$M$-FM-DCSK system over a multipath Rayleigh fading channel, where (a) $N = 64$, (b) $N = 128$, $SF = 480$, $M = 4$, ${D_{{\rm{sr}}}} = 6{\rm{m}}$, and ${D_{{\rm{rd}}}} = 10{\rm{m}}$.}
\label{fig:fig4}
\end{figure}
\subsubsection{Scheme II} \label{sect:Scheme II}
Fig.~\ref{fig:fig4} plots the theoretical and simulated BER curves for the source and relay in scheme II over a multipath Rayleigh fading channel, where the parameters $M$, $SF$, ${D_{{\rm{sr}}}}$, and ${D_{{\rm{rd}}}}$ are set to $4$, $480$, $6\rm{m}$, and $10\rm{m}$, respectively. It can be seen that the theoretical BER curves agree well with the simulated ones regardless of $N$ or $n$. Moreover, the BER performance of both the source and the relay in scheme II improves as $n$ increases, which is similar to the observation in scheme I. Furthermore, the BER performance of the source is better than that of the relay. The main reason is that the Euclidean distance of the source is larger than that of the relay in the signal space. For instance, when $N=128$, the source has about $3.5~{\rm{dB}}$ performance gain over the relay at a BER of ${10^{ - 4}}$ over a multipath Rayleigh fading channel.

\begin{figure}[tbp]
\center
\subfigure[\hspace{-0.8cm}]{ \label{fig:subfig:5a}
\includegraphics[width=3.2in,height=2.56in]{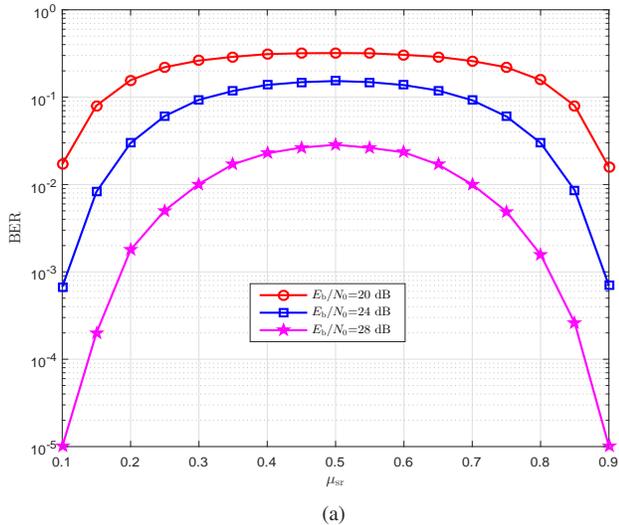}}
\subfigure[\hspace{-0.8cm}]{ \label{fig:subfig:5b}
\includegraphics[width=3.2in,height=2.56in]{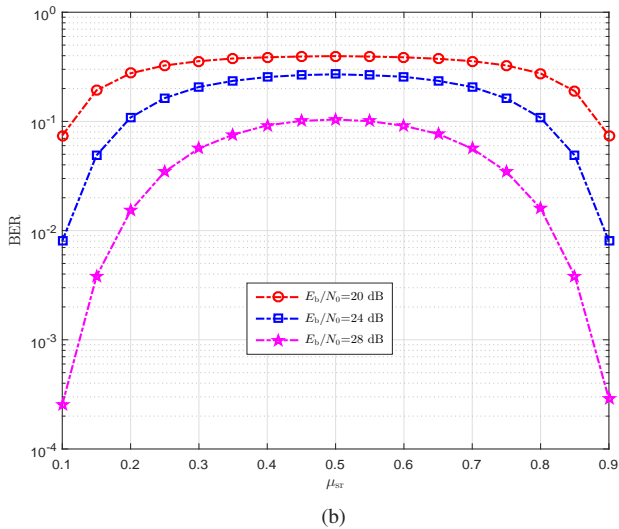}}
\caption{BER performance versus the parameter ${\mu _{{\rm{sr}}}}$  for the (a) source and (b) relay of scheme II in the RIS-$M$-FM-DCSK system over a multipath Rayleigh fading channel, where ${\mu _{{\rm{sr}}}} = \frac{{{D_{{\rm{sr}}}}}}{{{D_{{\rm{sr}}}} + {D_{{\rm{rd}}}}}}$, ${D_{{\rm{sr}}}} + {D_{{\rm{rd}}}} = 16{\rm{m}}$, $N = 64$, $SF = 480$, $M = 4$, and $n = 2$.}
\label{fig:fig5}
\vspace{-0.4cm}
\end{figure}
In Fig.~\ref{fig:fig5}, the BER curves for the source and relay are plotted against the parameter ${\mu _{{\rm{sr}}}} = \frac{{{D_{{\rm{sr}}}}}}{{{D_{{\rm{sr}}}} + {D_{{\rm{rd}}}}}}$, where total distance for the link source $ \to $relay$ \to $destination (i.e., ${D_{{\rm{sr}}}} + {D_{{\rm{rd}}}}$) is set to $16\rm{m}$.
We can find that when the relay is located at the position that has the same distance to the source and destination, both the source and relay have the worst BER performance.
The BER performance of both the source and relay gets better as the relay moves closer to the source or destination. The reason of this observation is related to the received SNR at the destination. Specifically, for a fixed ${E_{\rm{b}}}/{N_0}$, when ${\mu _{{\rm{sr}}}} = 0.5$, the destination has the smallest received SNR. Therefore, in scheme II, the relay should be deployed close to the source or destination to achieve desirable transmission reliability.

\begin{figure}[tbp]
\center
\includegraphics[width=3.2in,height=2.56in]{{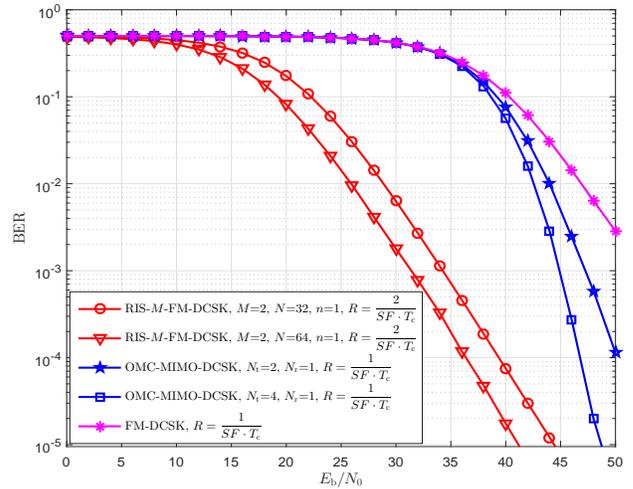}}
\vspace{-0.2cm}
\caption{BER performance of scheme I in the RIS-$M$-FM-DCSK system, the OMC-MIMO-DCSK system, and the FM-DCSK system over a multipath Rayleigh fading channel, where $SF = 360$ and ${D_{{\rm{sd}}}} = 16~{\rm{m}}$.}
\label{fig:fig6}  
\vspace{-2mm}
\end{figure}
\begin{figure}[tbp]
\center
\includegraphics[width=3.2in,height=2.56in]{{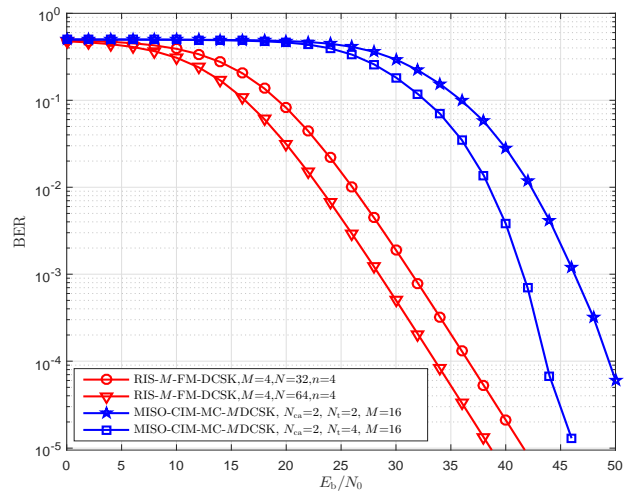}}
\vspace{-0.2cm}
\caption{BER performance of scheme I in the RIS-$M$-FM-DCSK system and the MISO-CIM-MC-$M$DCSK system over a multipath Rayleigh fading channel, where $SF = 360$ and ${D_{{\rm{sd}}}} = 16{\rm{m}}$.}
\label{fig:fig7}  
\vspace{-2mm}
\end{figure}
\subsection{BER Performance Comparison between the Proposed RIS-$M$-FM-DCSK System and Existing DCSK Counterparts} \label{sect:BER Performance Comparison between the Proposed RIS-$M$-FM-DCSK System and Existing Counterparts}

\subsubsection{Scheme I} \label{sect:SScheme I}
To better illustrate the superiority of the proposed RIS-$M$-FM-DCSK system, we compare the BER performance of the proposed system with other existing DCSK counterparts. Fig.~\ref{fig:fig6} shows the BER performance of scheme I, the FM-DCSK\cite{899922}, and the OMC-MIMO-DCSK system\cite{6661866} over a multipath Rayleigh fading channel, where $SF$ and ${D_{{\rm{sd}}}}$ are set to 360 and $16~\rm{m}$, respectively. As seen, the data rates $R$ are set to $\frac{2}{{SF \cdot {T_{\rm{c}}}}}$, $\frac{1}{{SF \cdot {T_{\rm{c}}}}}$, and $\frac{1}{{SF \cdot {T_{\rm{c}}}}}$ for the proposed scheme I ($M/n = 2/1$), the OMC-MIMO-DCSK, and the FM-DCSK systems, respectively. Note that $R = \frac{2}{{SF \cdot {T_{\rm{c}}}}}$ is the lowest achievable data rate for scheme I, whereas the OMC-MIMO-DCSK and FM-DCSK systems have a fixed data rate $R = \frac{1}{{SF \cdot {T_{\rm{c}}}}}$. For the OMC-MIMO-DCSK system, the number of transmit antennas ${N_{\rm{t}}}$ is set to $2$ and $4$, and the number of receive antennas ${N_{\rm{r}}}$ is set to $1$. It can be seen that the proposed scheme I outperforms the OMC-MIMO-DCSK and FM-DCSK systems over a multipath Rayleigh fading channel. For example, the proposed scheme I with $M/n/N = 2/1/32$ has about $10~\rm{dB}$ and $7~\rm{dB}$ performance gains over the OMC-MIMO-DCSK system with ${N_{\rm{t}}}/{N_{\rm{r}}} = 2/1$ and ${N_{\rm{t}}}/{N_{\rm{r}}} = 4/1$, respectively, at a BER of ${10^{ - 4}}$ over a multipath Rayleigh fading channel.

Fig.~\ref{fig:fig7} depicts the BER results of the proposed scheme I and the MISO-CIM-MC-$M$DCSK system\cite{9184069} over a multipath Rayleigh fading channel, where $SF$ and ${D_{{\rm{sd}}}}$ are set to $360$ and $16\rm{m}$, respectively. For a fair comparison, the data rates of the proposed scheme I ($M/n = 4/4$) and MISO-CIM-MC-$M$DCSK system (${N_{{\rm{ca}}}}/M = 2/16$) are set to $R = \frac{9}{{SF \cdot {T_{\rm{c}}}}}$. As observed, the proposed scheme I has better BER performance than the MISO-CIM-MC-$M$DCSK system. For instance, the proposed scheme I with $M/n/N = 4/4/32$ has about $5$~dB performance gain with respect to the MISO-CIM-MC-$M$DCSK system with ${N_{{\rm{ca}}}}/M/{N_{\rm{t}}} = 2/16/4$ at a BER of ${10^{ - 4}}$ over a multipath Rayleigh fading channel.

{\em Remark 1:}
Although the existing OMC-MIMO-DCSK and MISO-CIM-MC-$M$DCSK systems could achieve better BER performance by increasing the number of antennas, this would significantly increase the cost and hardware complexity.
\subsubsection{Scheme II} \label{sect:SScheme II}
\begin{figure}[tbp]
\center
\subfigure[\hspace{-0.8cm}]{ \label{fig:subfig:8a}
\includegraphics[width=3.2in,height=2.56in]{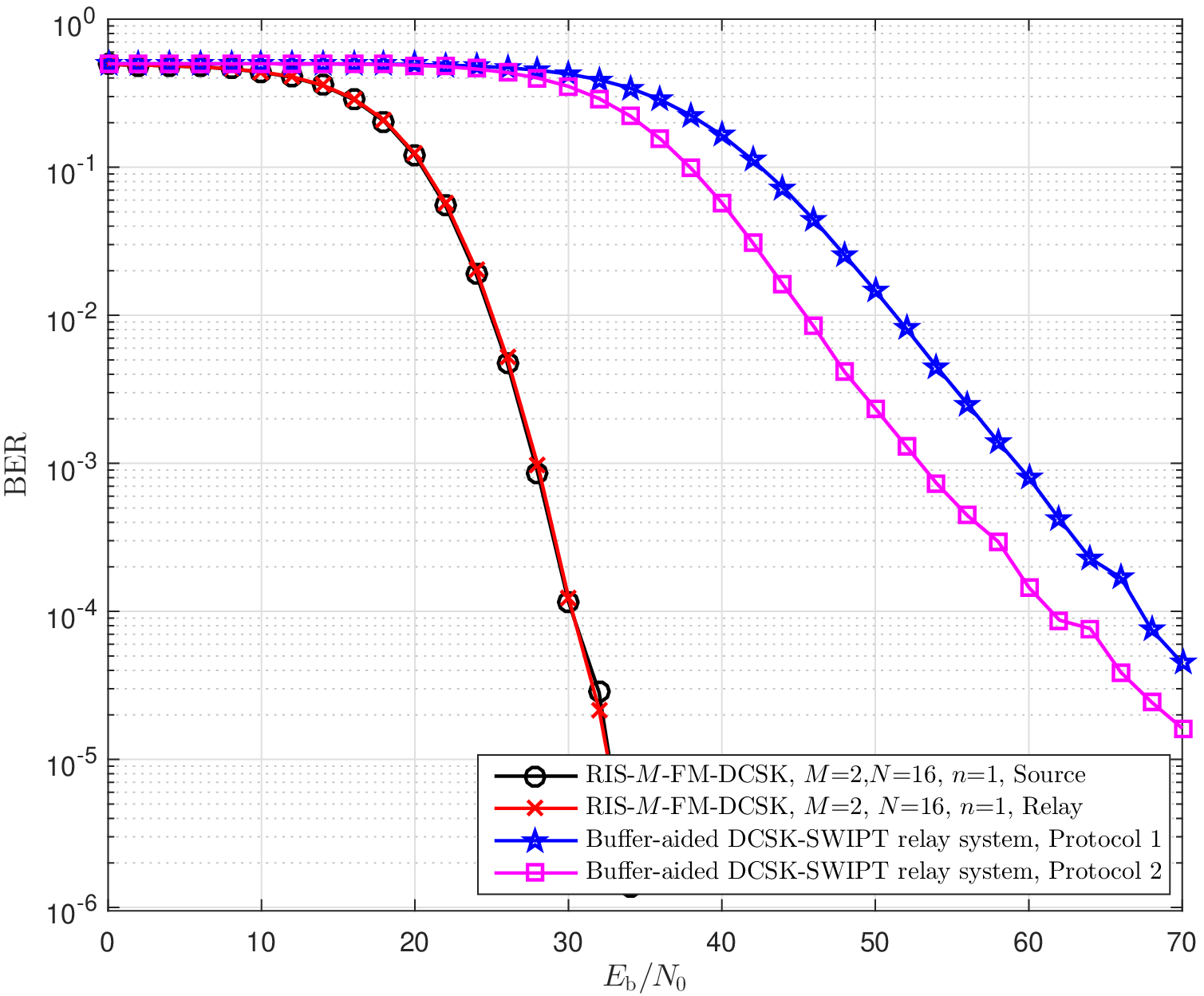}}
\subfigure[\hspace{-0.8cm}]{ \label{fig:subfig:8b}
\includegraphics[width=3.2in,height=2.56in]{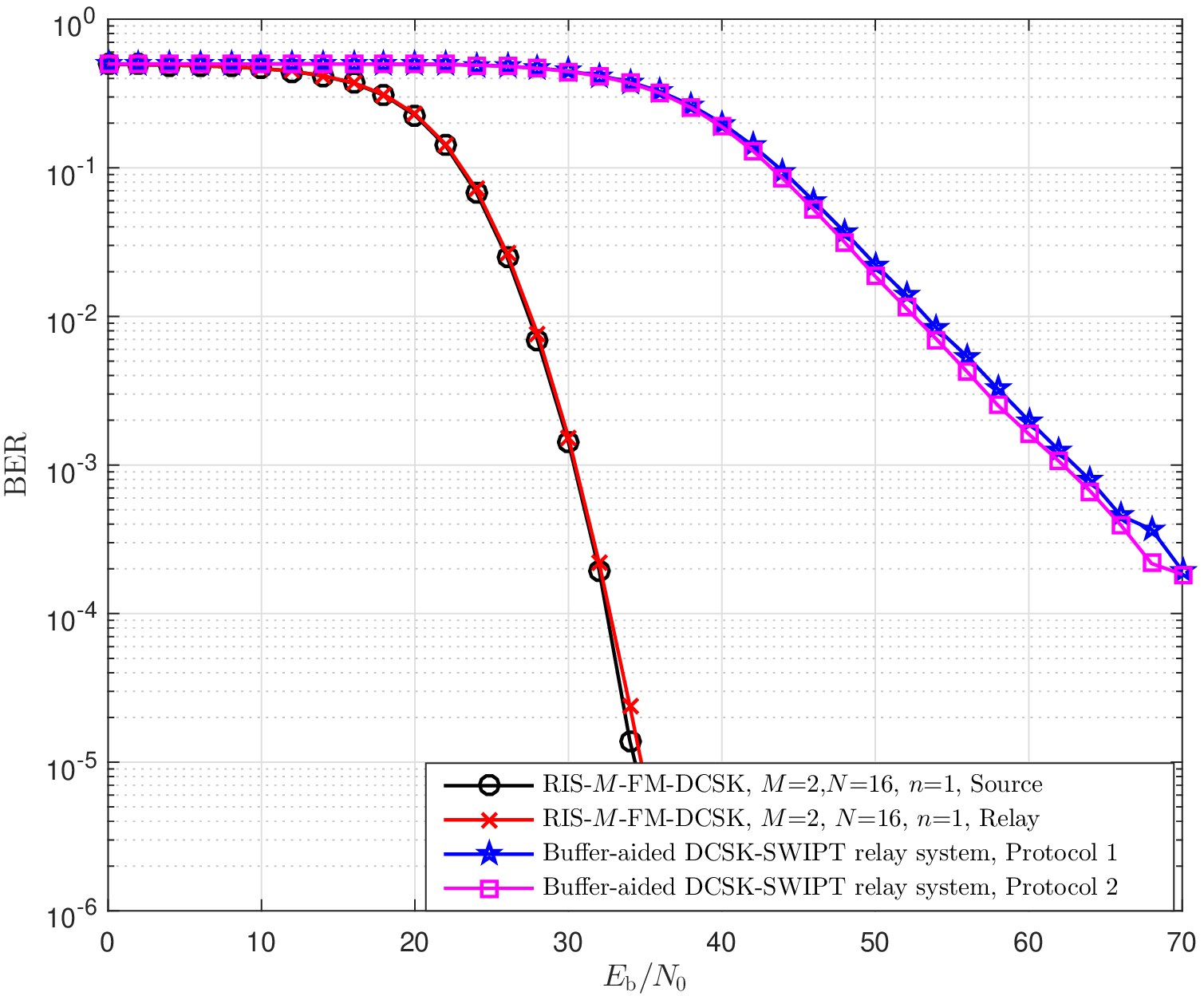}}
\caption{BER performance of scheme II in the RIS-$M$-FM-DCSK system and the buffer-aided DCSK-SWIPT relay system over a multipath Rayleigh fading channel, where $SF = 200$, (a) ${D_{{\rm{sr}}}} = 2{\rm{m}}$, ${D_{{\rm{rd}}}} = 6{\rm{m}}$, and (b) ${D_{{\rm{sr}}}} = 4{\rm{m}}$, ${D_{{\rm{rd}}}} = 4{\rm{m}}$.}
\label{fig:fig8}
\vspace{-0.4cm}
\end{figure}

Considering the CS-ii, we compare the BER performance between the proposed scheme II and buffer-aided DCSK-SWIPT relay system\cite{9142258} over a multipath Rayleigh fading channel in Fig.~\ref{fig:fig8}. The relays in the proposed scheme II and buffer-aided DCSK-SWIPT relay system are passive. Moreover, the buffer sizes for the protocol 1 and protocol 2 in the buffer-aided DCSK-SWIPT relay system are set to $10$. In Fig.~\ref{fig:fig8}, the data rate $R$ for the source and relay in scheme II ($M/n = 2/1$) are set to $R = \frac{1}{{SF \cdot {T_{\rm{c}}}}}$. However, the data rate of the buffer-aided DCSK-SWIPT relay system is adaptive according to the channel condition, and its maximum achievable rate is $R = \frac{1}{{SF \cdot {T_{\rm{c}}}}}$. When ${D_{{\rm{sr}}}} = 2{\rm{m}}$ and ${D_{{\rm{rd}}}} = 6{\rm{m}}$, the BER results for the proposed scheme II and the buffer-aided DCSK-SWIPT relay system are shown in Fig.~\ref{fig:subfig:8a}. It is obvious that the BER performance of both the source and relay in scheme II is much better than that of the buffer-aided DCSK-SWIPT relay system. For instance, both the source and relay in scheme II with $M/n = 2/1$  have about $30~\rm{dB}$ and $37~\rm{dB}$ performance gains over the protocol 1 and protocol 2 in the buffer-aided DCSK-SWIPT relay system, respectively, at a BER of ${10^{ - 4}}$ over a multipath Rayleigh fading channel. Besides, one can observe that the source and relay in scheme II with $M/n = 2/1$  have similar BER performance.

In the case of ${D_{{\rm{sr}}}} = 4{\rm{m}}$ and ${D_{{\rm{rd}}}} = 4{\rm{m}}$ (i.e., the relay is located at the same distance from both the source and destination), the BER performance of the proposed scheme II and the buffer-aided DCSK-SWIPT relay system over a multipath Rayleigh fading channel is presented in Fig.~\ref{fig:subfig:8b}. Although the deployment location of the relay is the most unfavorable for the proposed scheme II (see the discussion for Fig.~\ref{fig:fig5}), the BER performance of the proposed scheme II is still much better than that of the buffer-aided DCSK-SWIPT relay system. For instance, both the source and the relay in scheme II with $M/n = 2/1$ have about $30~\rm{dB}$ and $31~\rm{dB}$ performance gains with respect to the protocol 1 and protocol 2 in the buffer-aided DCSK-SWIPT relay system, respectively, at a BER of ${10^{ - 3}}$ over a multipath Rayleigh fading channel.

\section{Conclusion}\label{sect:Conclusions}
In this paper, we have investigated the design and performance analysis of a novel RIS-$M$-FM-DCSK system for realizing high-reliability transmissions in wireless environments. To achieve this objective, we have conceived two efficient transmission schemes for the proposed RIS-$M$-FM-DCSK system, referred to as scheme I and scheme II. Specifically, in scheme I, the RIS is treated as a transmitter at the source. Moreover, the information bits are carried by both the positive/negative state of the FM chaotic signal and the $M$-PSK symbols, which are obtained by the phase adjustment of the RIS. Thereby, scheme I is suitable for the CS that the direct link exists between the source and the destination. On the contrary, scheme II is suitable for the CS that the direct link between the source and the destination is blocked by an obstacle, such that the RIS is treated as a relay. In scheme II, the information bits of the source and relay can be sent to the destination simultaneously. In particular, the information bits of the source and relay are carried by the positive/negative state of the FM chaotic signal and the $M$-PSK symbols generated by the RIS, respectively. Furthermore, the theoretical BER expressions of the proposed RIS-$M$-FM-DCSK system over multipath Rayleigh fading channels have been derived. Simulated BER results match well with the corresponding theoretical derivations, demonstrating the accuracy of the proposed theoretical expressions. In addition, the proposed RIS-$M$-FM-DCSK system significantly outperforms the FM-DCSK, OMC-MIMO-DCSK, MISO-CIM-MC-$M$DCSK, and buffer-aided DCSK-SWIPT relay systems in terms of BER. As a consequence, the proposed RIS-$M$-FM-DCSK system can be considered as a promising candidate for low-power and low-cost wireless communication networks.

\bibliographystyle{IEEEtran}  
\bibliography{IEEEabrv,ref}

\end{document}